\documentclass[accepted]{uai2023} 

\usepackage[american]{babel}

\usepackage{natbib} 
\usepackage{balance}

\usepackage{mathtools} 
\usepackage{amsmath}
\usepackage{amssymb}
\usepackage{mathtools}
\usepackage{amsthm}

\theoremstyle{plain}
\newtheorem{theorem}{Theorem}[section]

\newtheorem{lemma}[theorem]{Lemma}

\theoremstyle{definition}
\newtheorem{definition}[theorem]{Definition}

\theoremstyle{remark}
\newtheorem{remark}[theorem]{Remark}

\usepackage[ruled,vlined]{algorithm2e}
\usepackage{booktabs} 
\usepackage{tikz} 

\usepackage{multirow}

\def\de{\overset{\Delta}{=}}

\SetKwInput{KwInput}{Input}
\SetKwInput{KwOutput}{Output}
\SetKwComment{Comment}{/*}{ */}
\let\oldnl\nl
\newcommand{\nonl}{\renewcommand{\nl}{\let\nl\oldnl}}

\title{Robust Quickest Change Detection for Unnormalized Models}

\author[1]{Suya Wu}
\author[1]{Enmao Diao}
\author[2]{Taposh Banerjee}
\author[3]{Jie Ding}
\author[1]{Vahid Tarokh}
\affil[1]{%
    Department of Electrical and Computer Engineering\\
    Duke University\\
    Durham, NC 27708, USA
}
\affil[2]{%
    Department of Industrial Engineering\\
    University of Pittsburgh\\
    Pittsburgh, PA 15213, USA
}
\affil[3]{%
    School of Statistics\\
    University of Minnesota Twin Cities\\
    Minneapolis, 
MN 55455, USA
  }
  
\begin{document}
\maketitle
\begin{abstract}
      Detecting an abrupt and persistent change in the underlying distribution of online data streams is an important problem in many applications. This paper proposes a new robust score-based algorithm called RSCUSUM, which can be applied to unnormalized models and addresses the issue of unknown post-change distributions. RSCUSUM replaces the Kullback-Leibler divergence with the Fisher divergence between pre- and post-change distributions for computational efficiency in unnormalized statistical models and introduces a notion of the ``least favorable'' distribution for robust change detection. The algorithm and its theoretical analysis are demonstrated through simulation studies. 
\end{abstract}

\section{Introduction}\label{sec: intro}
In the problem of quickest change detection, the objective is to detect an abrupt change in the 
statistical properties of an observed stochastic process. This change in the distribution has to be detected with the minimum possible delay, subject to a constraint on the rate of false alarms. This problem has applications in sensor networks, cyber-physical systems, biology, and neuroscience; see \cite{veeravalli2014quickest,  basseville1993detection, poor2008quickest, tartakovsky2014sequential}. 

When the pre- and post-change distribution of the data is known, a typical optimal algorithm in the literature is a stopping rule. A sequence of statistics is calculated using the likelihood ratio of the observations, and a change is declared when the sequence of statistics crosses a pre-designed threshold. The threshold is chosen to meet a constraint on false alarms; see \cite{shiryaev1963optimum,lorden1971procedures,pollak1985optimal,moustakides1986optimal,lai1998information,tartakovsky2005general}. The three most important algorithms in the literature are the Shiryaev algorithm (\cite{shiryaev1963optimum, tartakovsky2005general}), the cumulative sum (CUSUM) algorithm (\cite{page1955test, lorden1971procedures, moustakides1986optimal, lai1998information}), and the Shiryaev-Roberts algorithm (\cite{roberts1966comparison, pollak1985optimal}). 

The main challenge in implementing a change detection algorithm in practice is that the pre- and post-change distributions are not precisely known. This challenge is amplified when the data is high-dimensional. Specifically, in several machine learning applications, the data models may not lend themselves to explicit distributions. For example, energy-based models~(\cite{LeCun2006ATO}) capture dependencies between observed and latent variables based on their associated energy (an unnormalized probability), and score-based deep generative models~\cite{song2020score} generate high-quality images by learning the score function (the gradient of the log density function). These models can be computationally cumbersome to normalize themselves as probabilistic density functions. Thus, optimal algorithms from the change detection literature, which are likelihood ratio-based tests, are computationally expensive to implement. 

This issue is partially addressed in \cite{wuetal-aistat-2023} where the authors have proposed the SCUSUM algorithm, a Hyv\"arinen score-based (\cite{hyvarinen2005estimation}) modification of the CUSUM algorithm for quickest change detection. 
It is shown in \cite{wuetal-aistat-2023} that the SCUSUM algorithm is consistent and the authors also provide expressions for the average detection delay and the mean time to a false alarm. The 
Hyv\"arinen score is invariant to scale and hence can be applied to unnormalized models. This makes the SCUSUM algorithm highly efficient as compared to the classical CUSUM algorithm for high-dimensional models.

The main drawback of the SCUSUM algorithm is that its effectiveness is contingent on knowing the precise post-change unnormalized model, i.e., knowing the post-change model within a normalizing constant. In practice, due to a limited amount of training data, the post-change model can only be learned within an uncertainty class. To detect the change effectively, an algorithm must be robust against these modeling uncertainties. The SCUSUM algorithm is not robust in this sense. Specifically, if not carefully designed, the SCUSUM algorithm can fail to detect several (in fact, infinitely many) post-change scenarios. 

In this paper, we propose a robust score-based variant of the CUSUM algorithm for the quickest change detection. We refer to our algorithm as the RSCUSUM algorithm. Under the assumption that the post-change uncertainty class is convex and compact, we show that the RSCUSUM algorithm is robust, i.e., can consistently detect changes for every possible post-change model. This consistency is achieved by designing the RSCUSUM algorithm using the \textit{least favorable} distribution from the post-change class. 

The problem of optimal robust quickest change detection is studied in \cite{unnikrishnan2011minimax}. In a minimax setting, the optimal algorithm is the CUSUM algorithm designed using the least favorable distribution. 
The robust CUSUM test in \cite{unnikrishnan2011minimax} may suffer from two drawbacks: 1) It is a likelihood ratio-based test and hence may not be amenable to implementation in high-dimensional models. 2) The notion of least favorable distribution is defined using \textit{stochastic boundedness}, which may be difficult to verify for high-dimensional data. 

In contrast with the work in \cite{unnikrishnan2011minimax}, we define the notion of least favorable distribution using Fisher divergence and provide a method to effectively identify the least favorable distribution for the post-change model.

\subsection{Our Contributions}

We now summarize our contributions in this paper. 

$\bullet$ We propose a new robust score-based quickest change detection algorithm that can be applied to unnormalized models, namely, statistical models whose density involves an unknown normalizing constant. Specifically, we use the Hyv\"arinen score (\cite{hyvarinen2005estimation}) to propose a robust score-based variant of the SCUSUM algorithm from \cite{wuetal-aistat-2023}, which we refer to as RSCUSUM. In this variant and its subsequent theory, the role of Kullback-Leibler divergence in classical change detection is replaced with the Fisher divergence between the pre-and post-change distributions. Please see Section~\ref{sec:RSCUSUM_algorithm}.

$\bullet$ Our developed RSCUSUM algorithm can address unknown post-change models. Specifically, assuming that the post-change law belongs to a known family of distributions that is convex and compact, we identify a least favorable distribution that is closest in terms of Fisher divergence from the pre-change family. We then show that the RSCUSUM algorithm can consistently detect each post-change distribution from the family, and is robust in this sense. Please see Section~\ref{sec:theoritical_analysis}.

$\bullet$ 
We provide an effective method to identify the least favorable post-change distribution in a post-change family. This is in contrast to the setup in \cite{unnikrishnan2011minimax} where a stochastic boundedness characterization makes it harder to identify the least favorable distribution. Please see Section~\ref{sec:least_favorable_distribution}.

$\bullet$ 
From a theoretical perspective, unlike the CUSUM algorithm that leverages the fact that the likelihood ratios form a martingale under the pre-change model~\cite{lai1998information,woodroofe1982nonlinear}, the RSCUSUM algorithm is a score-based algorithm where cumulative scores do not enjoy a standard martingale characterization. Our analysis of the delay and false alarm analysis for RSCUSUM is based on new analysis techniques. Pleas see Section~\ref{sec:theoritical_analysis}.

$\bullet$ We demonstrate the effectiveness of the RSCUSUM algorithm through simulation studies on Gaussian and Gauss-Bernoulli Restricted Boltzmann Machine (RBM) models. Please see Section~\ref{sec:results}. 

\section{Problem Formulation}\label{sec:background}

\label{subsec:problem_formulation}
\noindent Let $\{X_n\}_{n\geq 1}$ denote a sequence of independent random variables defined on the probability space $(\Omega, \mathcal{F}, P_\nu)$. Let $\mathcal{F}_n$ be the $\sigma-$algebra generated by random variables $X_1,\; X_2, \;\dots,\; X_n$, and let $\mathcal{F}=\sigma(\cup_{n\geq 1}\mathcal{F}_n)$ be the $\sigma-$algebra generated by the union of sub-$\sigma$-algebras. 
Under $P_\nu$, $X_1, \; X_2, \;\dots,\; X_{\nu-1}$ are {i.i.d.} according to a density $p_\infty$ and $X_{\nu}, \; X_{\nu+1},\; \dots$ are {i.i.d.} according to a density $p_1$. We think of $\nu$ as the change point, $p_\infty$ as the pre-change density, and $p_1$ as the post-change density. We use $\mathbb{E}_{\nu}$ and $\text{Var}_{\nu}$ to denote the expectation and the variance associated with the
measure $P_\nu$, respectively. Thus, $\nu$ is seen as an unknown constant and we have an entire family $\{P_\nu\}_{1 \leq \nu \leq \infty}$ of change-point models, one for each possible change point. We use $P_\infty$ to denote the measure under which there is no change, with $\mathbb{E}_\infty$ denoting the corresponding expectation.

A change detection algorithm is a stopping time $T$ with respect to the data stream $\{X_n\}_{n\geq 1}$:
$$
\{T \leq n\} \in \mathcal{F}_n.
$$
If $T\geq \nu$, we have made a \textit{delayed detection}; otherwise, a \textit{false alarm} has happened. 
Our goal is to find a stopping time $T$ to optimize the trade-off between well-defined metrics on delay and false alarm. 
We consider two minimax problem formulations to find the best stopping rule. 

To measure the detection performance of a stopping rule, we use the following minimax metric (\cite{lorden1971procedures}), the worst-case averaged detection delay (WADD):
\begin{equation*}
\mathcal{L}_{\texttt{WADD}}(T)\de \sup_{\nu\geq 1}\text{ess}\sup \mathbb{E}_{\nu}[(T-\nu+1)^{+}|\mathcal{F}_{\nu}],
\end{equation*}
where $(y)^{+}\de\max(y, 0)$ for any $y\in \mathbb{R}$. Here $\text{ess} \sup$ is the essential supremum, i.e., the supremum outside a set of measure zero. We also consider the version of minimax metric introduced in \citet{pollak1985optimal}, the worst conditional averaged detection delay (CADD):
\begin{equation*}
    \mathcal{L}_{\texttt{CADD}}(T)\de \sup_{\nu\geq 1}\mathbb{E}_{\nu}[T-\nu|T\geq \nu].
\end{equation*}
For false alarms, we consider the \textit{average running length} (ARL), which is defined as the mean time to false alarm:
$$
\text{ARL}\de \mathbb{E}_\infty[T].
$$

We now formulate a robust quickest change detection problem; see \cite{unnikrishnan2011minimax}. We assume that pre- and post-change distributions are not precisely known. However, each is known within an uncertainty class: 
\begin{equation*}
    \begin{split}
        P_\infty &\in \mathcal{G}_\infty \\
        P_1 &\in \mathcal{G}_1. 
    \end{split}
\end{equation*}
For simplicity, in this paper, we will assume that the pre-change class is a singleton:
$$
\mathcal{G}_\infty = \{P_\infty\}.
$$
Our proposed method can also be extended to the
case of composite $\mathcal{G}_\infty$. The objective is to find a stopping rule to solve the following problem:
\begin{equation}
    \label{eq:lorden}
    \min_T \;\sup_{P_1 \in \mathcal{G}_1}\mathcal{L}_{\texttt{WADD}}(T)\;
    \quad \text{subject to}\;\quad \mathbb{E}_{\infty}[T]\geq \gamma,
\end{equation}
 where $\gamma$ is a constraint on the ARL. The delay $\mathcal{L}_{\texttt{WADD}}$ in the above problem is a function of the true post-change law $P_1$ and should be designated as
 $
 \mathcal{L}_{\texttt{WADD}}^{P_1}.
 $ 
 We will, however, suppress this notation and simply refer to $\mathcal{L}_{\texttt{WADD}}^{P_1}$ by $\mathcal{L}_{\texttt{WADD}}$. 
 Thus, the goal in this problem is to find a stopping time $T$ to minimize the worst-case detection delay, subject to a constraint $\gamma$ on $\mathbb{E}_{\infty}[T]$. 
 
We are also interested in the version with the minimax metric introduced in \citet{pollak1985optimal}: 
\begin{equation}
    \label{eq:pollak}
    \min_T \;\sup_{P_1 \in \mathcal{G}_1} \mathcal{L}_{\texttt{CADD}}(T)\;
    \quad \text{subject to}\;\quad \mathbb{E}_{\infty}[T]\geq \gamma. 
\end{equation}
If the post-change family is also singleton, $\mathcal{G}_1 = \{P_1\}$, then the above formulations are the classical minimax formulations from the quickest change detection literature; see \cite{veeravalli2014quickest, tartakovsky2014sequential, poor2008quickest}. The optimal algorithm (exactly optimal for \eqref{eq:lorden} and asymptotically optimal for \eqref{eq:pollak}) is the CUSUM algorithm given by
\begin{equation*} 
    T_{\texttt{CUSUM}}=\inf\{n\geq 1:\Lambda(n)\geq \tau\},
\end{equation*}
where $\Lambda(n)$ is defined using the recursion
\begin{align}
    &\Lambda(0)=0, \nonumber \\
    &\Lambda(n) \de \biggr(\Lambda(n-1)+\log \frac{p_1(X_n)}{p_{\infty}(X_n)}\biggr)^{+}, \forall n \geq 1, \label{eq:cusum_score}
 \end{align}
which leads to a computationally convenient stopping scheme. We recall that here $p_1$ is the post-change density and $p_\infty$ is the pre-change density.

In \cite{lorden1971procedures} and \cite{lai1998information}, the asymptotic performance of the CUSUM algorithm is also characterized. Specifically, it is shown as $\gamma \rightarrow \infty$,
\begin{align*}
    \mathcal{L_{\texttt{WADD}}}(T_{\texttt{CUSUM}}) \sim \mathcal{L_{\texttt{CADD}}}(T_{\texttt{CUSUM}})\sim \frac{\log \gamma}{\mathbb{D}_{\texttt{KL}}(P_{1}\|P_{\infty})}.
\end{align*}
Here $\mathbb{D}_{\texttt{KL}}(P_{1}\|P_{\infty})$ is the Kullback-Leibler divergence between the post-change distribution and pre-change distribution:
$$
\mathbb{D}_{\texttt{KL}}(P_{1}\|P_{\infty}) = \int p_1(x) \log \frac{p_1(x)}{p_\infty(x)} dx, 
$$
and the notation $g(c)\sim h(c)$ as $c\to c_0$ indicates that $\frac{g(c)}{h(c)} \to 1$ as $c\to c_0$ for any two functions $c\mapsto g(c)$ and $c\mapsto h(c)$.

Since the CUSUM algorithm uses likelihood ratio to compute its statistic, it is not amenable to implementation for high-dimensional models (see \cite{wuetal-aistat-2023}), where often the densities $p_1$ or $p_\infty$ are only known within a normalizing constant.

\section{Robust Quickest Change Detection for Unnormalized Models} \label{sec:RSCUSUM_algorithm}
\noindent In this section, we propose a robust score-based CUSUM (RSCUSUM) algorithm. We first review the SCUCUM algorithm proposed by \cite{wuetal-aistat-2023} to address the issues with likelihood ratio-based CUSUM for unnormalized models. The SCUSUM is defined based on Hyv\"arinen Score (\cite{hyvarinen2005estimation}), which circumvents the computation issue of the normalization constant. Similar to the schemes of SCUSUM, we use the Hyv\"arinen score and propose a robust variant that releases the knowledge of the true post-change distribution, where we assume the true post-change distribution is unknown but its uncertainty class is known.

Recall from Section~\ref{subsec:problem_formulation} that under the measure $P_\infty$, there is no change, and the density for each random variable is $p_\infty$. In the rest of the paper, we refer to the probability measure of $X_1$ under $P_\infty$, also by $P_\infty$. Similarly, we refer to the law of $X_1$ under $P_1$ also by $P_1$. The differences will always be clear from the context. 

We provide the definition of the Hyv\"arinen Score below.
\begin{definition}[Hyv\"arinen Score] The Hyv\"arinen score of any measure $P$ (with density $p$) is a mapping $(X, P)\mapsto \mathcal{S}_{\texttt{H}}(X, P)$ given by 
    \begin{equation*}
        \mathcal{S}_{\texttt{H}}(X, P) \de \frac{1}{2} \left \| \nabla_{X} \log p(X) \right \|_2^2 + \Delta_{X} \log p(X)
    \end{equation*}
 whenever it can be well defined. Here, $\|\cdot\|_2$ denotes the Euclidean norm, $\nabla_{X}$ and $\Delta_{X} = \sum_{i=1}^d \frac{\partial^2}{\partial x_i^2}$ respectively denote the gradient and the Laplacian operators acting on $X = (x_1, \cdots, x_d)^{\top}$.
\end{definition}

By using the Hyv\"arinen Score in our algorithm, the role of Kullback-Leibler divergence in the theoretical analysis of the algorithm is replaced by the Fisher divergence. 

\begin{definition}[Fisher Divergence] The Fisher divergence between two probability measures $P$ to $Q$ (with densities $p$ and $q$) is defined by
\begin{align*}
    \mathbb{D}_{\texttt{F}} (P \| Q) \de \mathbb{E}_{X\sim P} \left[\left \| \nabla_{{X}} \log p(X)- \nabla_{{X}} \log q(X)\right \|_2^2 \right],
\end{align*}
whenever the integral is well defined. 
\end{definition}

Clearly, $\nabla_{{X}} \log p(X)$, $\nabla_{{X}} \log q(X)$, and $\Delta_{X} \log q(X)$ remain invariant if $p$ and $q$ are scaled by any positive constant with respect to $X$. Hence, the Fisher divergence and the Hyv\"arinen Score remain \textit{scale-variant} concerning an arbitrary constant scaling of density functions.

The SCUSUM~\citep{wuetal-aistat-2023} assumes that the true pre- and post-chagne distributions $P_{\infty}$ and $P_{1}$ are known. It defines the detection score by
\begin{equation}
    \tilde{z}_{\lambda}(X)\de \lambda\bigr(\mathcal{S}_{\texttt{H}}(X, P_{\infty})-\mathcal{S}_{\texttt{H}}(X, P_{1})\bigr).
\end{equation}
However, it is impractical, in particular for online data streams, to know the true post-change distribution. We assume that pre-change data is available. This data and a model class $\mathcal{G}_\infty$ are used to model/learn the pre-change distribution $P_{\infty}$. The post-change distribution $P_{1}$ is assumed to be modeled by an unknown element of a parametric family $\mathcal{G}_1 = \{G_{\theta}:\,\theta \in \Theta_1\}$. We note that our framework readily extends to the case of non-parametric families but for simplicity, we present our results only in the parametric case.

We define the notion of least favorable distribution. This approach to defining the least favorable distribution for the quickest change detection is novel. 
\begin{definition}[Least Favorable Distribution (LFD)]
\label{defination: LFD}
    Assume that the family $\mathcal{G}_1 = \{G_{\theta} :\,\theta \in \Theta_1\}$ is convex and compact. We define
    \begin{equation}
\label{eq:Q1LFD}
   Q_{1}  = \arg \min_{ G_{\theta} \in \mathcal{G}_1} \mathbb{D}_{F}( G_{\theta} \| P_{\infty}).
\end{equation}
\end{definition}

The existence of $Q_{1}$ is guaranteed by the compactness of $\mathcal{G}_1$ and the continuity of the Fisher divergence as a function of its arguments. Thus, $Q_1$ is the closest element of $\mathcal{G}_1$ to $P_\infty$ in the Fisher-divergence sense.

Given the pre-change law $P_\infty$ (with density $p_\infty$), we now use $Q_1$ and its density $q_1$ to design the RSCUSUM algorithm. 
We define the instantaneous RSCUSUM score function $X\mapsto z_{\lambda}(X)$ by 
\begin{equation}
\label{eq:scusum_instantaneous}
    z_{\lambda}(X) \de \lambda\bigr(\mathcal{S}_{\texttt{H}}(X, P_{\infty})-\mathcal{S}_{\texttt{H}}(X, Q_{1})\bigr),
\end{equation}
where $\lambda>0$ is a pre-selected multiplier, $\mathcal{S}_{\texttt{H}}(X, P_{\infty})$ and $\mathcal{S}_{\texttt{H}}(X, Q_1)$ are respectively the Hyv\"arinen score functions of $P_\infty$ and $Q_1$. If the post-change model is precisely known, then the $Q_1$ in the above equation will be replaced by the known post-change law and RSCUSUM is identical to SCUSUM~\citep{wuetal-aistat-2023}. In Section~\ref{sec:theoritical_analysis}, we will provide more discussion on the role of $\lambda$ in the RSCUSUM algorithm. 

Our proposed stopping rule is given by 
\begin{equation} 
\label{eq:SCUSUM_rule}
    T_{\texttt{RSCUSUM}}=\inf\{n\geq 1:Z(n)\geq \tau\},
\end{equation}
where $\tau>0$ is a stopping threshold that is pre-selected to control false alarms, and $Z(n)$ can be computed recursively:
\begin{align*}
    &Z(0)=0, \\
    &Z(n) \de (Z(n-1)+z_{\lambda}(X_n))^{+},\;\forall n\geq 1.
\end{align*}
The statistic $Z(n)$ is referred to as the detection score of RSCUSUM at time $n$. The RSCUSUM algorithm is summarized in Algorithm~\ref{algm:rscusum}.


\begin{algorithm}
\DontPrintSemicolon
\caption{RSCUSUM Detection Algorithm}
\label{algm:rscusum}
\KwInput{Hyvarinen score functions $\mathcal{S}_{\texttt{H}}(\cdot, P_{\infty})$ and $\mathcal{S}_{\texttt{H}}(\cdot, Q_{1})$ of pre-change distribution and least favorable distribution in $\mathcal{G}_1$, respectively.} 
\KwData{$m$ previous observations $\mathbf{X}_{[-m+1,0]}$ and the online data stream $\{X_n\}_{n\geq 1}$}
\SetKwProg{Fn}{Initialization}{:}{}
  \Fn{}{
       Current time $k=0$, $\lambda>0$, $\tau>0$, and $Z(0)=0$}
\While{$Z(k)<\tau$}{
$k = k+1$\\
Update $z_{\lambda}(X_k) = \lambda(\mathcal{S}_{\texttt{H}}(X_{k}, P_{\infty})-\mathcal{S}_{\texttt{H}}(X_{k}, Q_{1}))$\\
Update $Z(k) = \max(Z(k-1)+z_{\lambda}(X_k), 0)$\;
}
Record the current time $k$ as the stopping time $T_{\texttt{RSCUSUM}}$\;
\KwOutput{$T_{\texttt{RSCUSUM}}$}
\end{algorithm}

\section{Delay and False Alarm Analysis of the RSCUSUM Algorithm} \label{sec:theoritical_analysis}
\noindent In this section, we provide delay and false alarm analysis of the RSCUSUM algorithm. 
We introduce two assumptions: 1) $P_{\infty}\notin\mathcal{G}_1$, and 2) the same mild regularity conditions introduced in \cite{hyvarinen2005estimation} so that the Hyv\"arinen score is well-defined.

We first prove an important lemma for our problem. If the Fisher divergence is seen as a measure of distance between two probability measures, then the following lemma provides a reverse triangle inequality for this distance, under the mild assumption that the order of integrals and derivatives can be interchanged.
\begin{lemma}
\label{lemma:tech}
Let $P_{\infty}$ be the pre-change distribution, $Q_1\in \mathcal{G}_1$ be the least-favorable distribution (as defined in Equation~\ref{eq:Q1LFD}), and $Q_2 \in \mathcal{G}_1$ be any other post-change distribution. Then 
\begin{equation*}
\mathbb{D}_{\texttt{F}}\left(Q_1 \| P_{\infty}\right)\leq \mathbb{D}_{\texttt{F}}\left(Q_2 \| P_{\infty}\right) - \mathbb{D}_{\texttt{F}}\left(Q_2 \| Q_1\right).
\end{equation*}
\end{lemma}
\begin{proof}
Consider a convex set of densities \begin{align*}
\bigl\{y\mapsto q_{\xi}(x): q_{\xi}(x)=\xi q_1(x)+(1-\xi) q_2(x), \xi \in [0,1]\bigr\},
\end{align*}
where $q_1$ and $q_2$ are densities of $Q_1$ and $Q_2$, respectively. 
Let $Q_{\xi}$ denote the distribution characterized by density $q_{\xi}$. 
We note that $Q_{\xi} \in \mathcal{G}_1$ due to the convexity assumption on $\mathcal{G}_1$. We use $\mathcal{L}(\xi)$ to denote the Fisher divergence $\mathbb{D}_{\texttt{F}} \left(Q_{\xi}\| P_{\infty}\right)$, and
\begin{align*}
    \mathcal{L}(\xi)&=\int \big\|\nabla \log q_{\xi}-\nabla\log p_{\infty}\big\|^2 q_{\xi} dx\\
    &=\int \big\| \nabla \log \bigl(\xi q_1+(1-\xi)q_2\bigr)-\nabla \log p_{\infty} \big\|^2\\
    &\qquad \qquad \qquad \qquad \qquad \bigl(\xi q_1+(1-\xi)q_2\bigr) dx.
\end{align*}
Clearly $\mathcal{L}(\xi)$ is minimized at $\xi=1$, and $\frac{\partial \mathcal{L}(\xi)}{\partial\xi}\mid_{\xi=1^-}\le 0$. 
Let $\mathcal{L}^{\prime}(\xi)=\frac{\partial \mathcal{L}(\xi)}{\partial\xi}$, we have 
\begin{align*}
&\mathcal{L}^{\prime}(\xi)=\int\bigl(q_1-q_2\bigr)\big\|\nabla \log q_{\xi}-\nabla \log p_{\infty}\big\|^2 d x\\
& \quad+\int 2q_{\xi} \nabla\left( \frac{q_1-q_2}{q_{\xi}} \right)^T \bigl(\nabla\log q_{\xi} -\nabla \log p_{\infty}\bigr)dx.
\end{align*}
This implies 
\begin{align}
\label{eq:diff}
&\mathcal{L}^{\prime}(1^{-})= \int\bigl(q_1-q_2\bigr)\big\|\nabla \log q_1- \nabla \log p_{\infty}\big\|^2 dx\notag\\
    & \quad +\int 2q_1 \nabla\left(\frac{q_1-q_2}{q_1}\right)^T\bigl(\nabla \log q_1-\nabla \log p_{\infty}\bigr) dx\notag\\
&= \mathbb{D}_{\texttt{F}}\left(Q_1 \| P_{\infty}\right)-\int\underbrace{ q_2\big\|\nabla\log q_1-\nabla \log p_{\infty}\big\|^2}_{\text{term 1}}\notag\\
    &\qquad \quad+\underbrace{2 q_1 \nabla\left(\frac{q_1-q_2}{q_1}\right)^T \bigl(\nabla \log q_1-\nabla\log p_{\infty}\bigr)}_{\text{term 2}}dx.
\end{align}
For term 1, we have 
\begin{align}
\label{eq:term1}
&q_2\big\| \nabla \log q_1-\nabla\log p_{\infty}\big\|^2 \nonumber\\
&=q_2\big\|\nabla\log q_1-\nabla\log q_2\big \|^2+q_2\big\|\nabla\log q_2-\nabla\log p_{\infty}\big\|^2\nonumber\\
&\quad+\underbrace{2q_2\bigl(\nabla \log q_1-\nabla\log q_2\bigr)^T \bigl(\nabla \log q_2-\nabla\log p_{\infty}\bigr)}_{\text{term 1(a)}}.
\end{align}
We note that,
\begin{align}
    &\int_x q_2
\big\|\nabla\log q_1-\nabla\log q_2\big\|^2 dx = \mathbb{D}_{\textit{F}}(Q_2\|Q_1), \label{eq: fisher1}\\
& \int_x q_2
\big\|\nabla\log q_2-\nabla\log p_{\infty}\big\|^2 dx = \mathbb{D}_{\textit{F}}(Q_2\|P_{\infty}). \label{eq: fisher2}
\end{align}
For term 2, we note that 
\begin{align*}
\nabla\left( \frac{q_1-q_2}{q_1}\right)
=\frac{q_2}{q_1}\bigl(\nabla\log q_1-\nabla\log q_2\bigr).
\end{align*}
Therefore, 
\begin{align}
&2 q_1 \nabla\left(\frac{q_1-q_2}{q_1}\right)^T \bigl(\nabla \log q_1-\nabla\log p_{\infty}\bigr)\nonumber\\
&=2q_2\bigl(\nabla \log q_1-\nabla\log q_2\bigr)^T \bigl(\nabla \log q_1-\nabla\log p_{\infty}\bigr).
\label{eq:term2}
\end{align}
Combining the last term in Equation (\ref{eq:term1}) with Equation (\ref{eq:term2}),
\begin{align}
\label{eq: comb_term12}
&-\text{term 1(a)} + \text{term 2} \notag\\
&=2q_2\bigl(\nabla \log q_1-\nabla\log q_2\bigr)^T \notag \\
    & \qquad \bigl(\nabla \log q_1-\nabla\log p_{\infty} -\nabla\log q_2 +\nabla\log p_{\infty} \bigr) \notag\\
&=2q_2\|\nabla \log q_1 - \nabla \log q_2\|^2.
\end{align}
Plugging Equations (\ref{eq: fisher1}), (\ref{eq: fisher2}), and (\ref{eq: comb_term12}) into Equation~(\ref{eq:diff}), 
\begin{align*}
&\mathcal{L}^{\prime}(1^{-})=\mathbb{D}_{\texttt{F}}\left(Q_1 \| P_{\infty}\right)+\mathbb{D}_{\texttt{F}}\left(Q_2 \| Q_1\right)-\mathbb{D}_{\texttt{F}}\left(Q_2 \| P_{\infty}\right).
\end{align*}
The results follows since $\frac{\partial \mathcal{L}(\xi)}{\partial \xi}\mid_{\xi=1^-}\le 0$.
\end{proof}

We now use Lemma~\ref{lemma:tech} to prove a result on the consistency of our proposed RSCUSUM algorithm.

\begin{lemma}[Positive and Negative Drifts]
\label{lemma: drifts}
Consider the instantaneous RSCUSUM score function $X\mapsto z_{\lambda}(X)$ as defined in Equation~(\ref{eq:scusum_instantaneous}). Recall that $P_1 \in \mathcal{G}_1$ is the true (but unknown) post distribution. Then,
\begin{align*}
\label{eq:expst}
    &\mathbb{E}_{\infty}\left[z_{\lambda}(X)\right] = -\lambda\mathbb{D}_{\texttt{F}}(P_{\infty} \| Q_1)<0,\; \text{and}\\
    &\mathbb{E}_{1}\left[z_{\lambda}(X)\right] 
 \ge \lambda\mathbb{D}_{\texttt{F}}(Q_1 \| P_{\infty})>0.
\end{align*}
\end{lemma}
\begin{proof}
Under some mild regularity conditions, \cite{hyvarinen2005estimation} proved that
\begin{align*}
    \mathbb{D}_{\texttt{F}} (P \| Q) =\mathbb{E}_{X\sim P} \left[\frac{1}{2}\left \| \nabla_{X} \log p(X) \right \|_2^2 + \mathcal{S}_{\texttt{H}}( X, Q)\right].
\end{align*}
We use $C_P$ to denote the term $\mathbb{E}_{X\sim P} \left[\frac{1}{2}\left \| \nabla_{X} \log p(X) \right \|_2^2\right]$. Then 
\begin{equation*}
\begin{split}
     \mathbb{E}_{\infty}&[\mathcal{S}_{\texttt{H}}(X, P_{\infty})-\mathcal{S}_{\texttt{H}}(X, Q_1)]\\
     &=\mathbb{D}_{\texttt{F}} (P_{\infty} \| P_{\infty})-C_{P_{\infty}}-\mathbb{D}_{\texttt{F}} (P_{\infty} \| Q_1)+C_{P_{\infty}}\\
     &=-\mathbb{D}_{\texttt{F}} (P_{\infty} \| Q_1),
     \end{split}
\end{equation*}
and 
\begin{equation*}
\begin{split}
     \mathbb{E}_{1}&[\mathcal{S}_{\texttt{H}}(X, P_{\infty})-\mathcal{S}_{\texttt{H}}(X, Q_1)]\\
     &=\mathbb{D}_{\texttt{F}} (P_1 \| P_{\infty})-C_{P_{1}}-\mathbb{D}_{\texttt{F}} (P_1 \| Q_1)+C_{P_{1}} \\
     &\ge \mathbb{D}_{\texttt{F}} (Q_1 \| P_{\infty}),
     \end{split}
\end{equation*}
where we applied Lemma \ref{lemma:tech}.

Since $\lambda>0$, the results follow.
\end{proof}

Lemma~\ref{lemma: drifts} shows that, prior to the change, the expected mean of instantaneous RSCUSUM score  $z_{\lambda}(X)$ is negative. Consequently, the accumulated score has a negative drift at each time $n$ prior to the change. Thus, the RSCUSUM detection score $Z(n)$ is pushed toward zero before the change point. This intuitively makes a false alarm unlikely. In contrast, after the change, the instantaneous score has a positive mean, and the accumulated score has a positive drift. Thus, the RSCUSUM detection score will increase toward infinity and leads to a change detection event.

Next, we discuss the values of the multiplier $\lambda$ in the theoretical analysis. Obviously, with a fixed stopping threshold, a larger value of $\lambda$ results in a smaller detection delay because the increment of the SCUSUM detection score is large, and the threshold can be easily reached. However, a larger value of $\lambda$ also causes SCUSUM to stop prematurely when no change occurs, leading to a larger false alarm probability. Hence, the value of $\lambda$ cannot be arbitrarily large (except in the degenerate case where $P_{\infty}(S_{\texttt{H}}(X, Q_1)-S_{\texttt{H}}(X, P_{\infty})\le 0)=1$). It needs to satisfy the following key condition:
\begin{equation}
\label{eq: condition}    
\mathbb{E}_{\infty}[\exp(z_{\lambda}(X))]\leq 1.
\end{equation}
We will present a technical lemma that guarantees the existence of such a $\lambda$ to satisfy inequality~(\ref{eq: condition}). 

\begin{lemma}[Existence of appropriate $\lambda$]
    \label{lemma: lambda}
There exists $\lambda>0$ such that Inequality~(\ref{eq: condition}) holds. Moreover, either 1) there exists $\lambda^{\star} \in (0,\infty)$ such that the equality of~(\ref{eq: condition}) holds, or 2) for all $\lambda>0$, the inequality of~(\ref{eq: condition}) is strict. As noted in \cite{wuetal-aistat-2023}, the second case is of no practical interest.
\end{lemma}
\begin{proof}
    We give proof in the supplementary material.
\end{proof}
From now on, we consider a fix $\lambda > 0$ that satisfies Inequality~(\ref{eq: condition}) to present our core results. In practice, it is possible to use $m$ past samples $\mathbf{X}_{[-m+1,0]}$ to determine the value of $\lambda$. In particular, $\lambda$ can be chosen as the positive root of the function $\lambda \to \tilde{h}(\lambda)$ given by 
\begin{align*}
\tilde{h}(\lambda)\de\frac{1}{m}\sum_{i=1}^m[\exp(z_{\lambda}(X_{i-m}))]-1.
\end{align*}
By Lemma~\ref{lemma: lambda} and its related technical discussions, the above equation has a root greater than zero with a high probability if $m$ is sufficiently large. In the case that $\lambda$ is not chosen properly, the algorithm remains implementable but optimal performance of detection delay is not guaranteed. We discuss this situation further in the supplementary material. 

\begin{theorem}
\label{thm:arl}
Consider the stopping rule $T_{\texttt{RSCUSUM}}$ defined in Equation~(\ref{eq:SCUSUM_rule}). Then, for any $\tau>0$,
    \begin{equation*}
        \mathbb{E}_{\infty}[T_{\texttt{RSCUSUM}}]\geq  e^{\tau}.
    \end{equation*}
    To satisfy the constraint of $\mathbb{E}_{\infty}[T_{\texttt{RSCUSUM}}] \geq \gamma$, it is enough to set the threshold $\tau=\log \gamma$. 
\end{theorem}
\begin{proof}
    We give proof in the supplementary material.
\end{proof}
Theorem~\ref{thm:arl} implies that the ARL increases at least exponentially as the stopping threshold increases. 

The following theorem gives the asymptotic performance of the RSCUSUM algorithm in terms of the detection delay under the control of the ARL.

\begin{theorem}
\label{thm:cond_edd}
   Subject to $\mathbb{E}_{\infty}[T_{\texttt{RSCUSUM}}]\geq \gamma>0$, the stopping rule $T_{\texttt{RSCUSUM}}$ satisfies
\begin{align*}
    \mathcal{L}_{\texttt{WADD}}&(T_{\texttt{RSCUSUM}}) \sim \mathcal{L}_{\texttt{CADD}}(T_{\texttt{RSCUSUM}}) \sim \mathbb{E}_1[T_{\texttt{RSCUSUM}}]\\
    &\sim \frac{\log \gamma}{\lambda (\mathbb{D}_{\texttt{F}}(P_1\|P_{\infty})-\mathbb{D}_{\texttt{F}}(P_1\|Q_{1}))}\\
    &\lesssim \frac{\log \gamma}{\lambda \mathbb{D}_{\texttt{F}}(Q_1 \| P_\infty)}, \quad \text{as $\gamma \to \infty$.  }
\end{align*} 
\end{theorem}
\begin{proof}
    We give proof in the supplementary material. 
\end{proof}

In the above theorem, we have used the notation $g(c)\lesssim h(c)$ as $c\to c_0$ to indicate that $\lim \sup \frac{g(c)}{h(c)} \leq 1$ as $c\to c_0$ for any two functions $c\mapsto g(c)$ and $c\mapsto h(c)$.

Theorems~\ref{thm:arl} and~\ref{thm:cond_edd} imply that the \textit{expected detection delay} (EDD) increases linearly as the stop threshold $\tau$ increases subject to a constraint on ARL.

\section{Identification of the least favorable distribution}
\label{sec:least_favorable_distribution}
Consider a general parametric distribution family $\mathcal{P}$ defined on $\mathcal{X}$. We use $\mathcal{P}_m$ to denote a set of a finite number of distributions belonging to $\mathcal{P}$, namely \begin{align*}
    \mathcal{P}_m = \{P_i,\; i=1,\dots, m:\; P_i\in \mathcal{P}\},\; m\in \mathbb{N}^{+}.
\end{align*}
We use $p_i$ to denote the density of each distribution $P_i, \;i=1, \dots, m$. Then, we define a convex set of densities 
    \begin{multline}
    \label{eq:convex_post_family}
        \mathcal{A}_m \de \biggl\{ x \mapsto \sum_{i=1}^m \alpha_i p_i(x):  \sum_{i=1}^m \alpha_i=1, \alpha_i \geq 0\biggr\}. 
    \end{multline}
We further define a set of functions
    \begin{multline}
    \label{eq:convex_gradient_density}
        \mathcal{B}_m \de \biggl\{ x \mapsto \sum_{i=1}^m \beta_i(x) \nabla_x \log  p_i(x): \\
         \sum_{i=1}^m \beta_i(x)=1,\; \beta_i(x) \geq 0,\; p_i \in \mathcal{P}_m \biggr\}. 
    \end{multline}
Consider the pre-change distribution $P_{\infty}$ (with density $p_{\infty}$) such that $P_{\infty} \in \mathcal{P}$ and $P_{\infty}\notin \mathcal{A}_m$. We use $\mathbb{E}_{\infty}$ to denote its corresponding expectation with $p_{\infty}$. Next, we provide a result to identify the LFD in $\mathcal{A}_m$ in terms of the Fisher-divergence (as defined in Definition~\ref{defination: LFD}). 
    
\begin{theorem} \label{thm_general_LFD}
    Assume that there exists an element $P_0 \in \mathcal{A}_m$  (with density $p_0$) such that
    \begin{multline}
      \mathbb{E}_{p_0}\biggl\{ \|\nabla_x \log p_0(X) -\nabla_x \log p_{\infty}(X) \|_2^2 \biggr\} \\
        = \min_{p \in \mathcal{A}_m, \phi \in \mathcal{B}_m} \mathbb{E}_{p} \biggl\{\|\phi (X) -\nabla_x \log p_{\infty}(X) \|_2^2 \biggr\}. 
        \label{eq10}
    \end{multline}
    Then, we have
    \begin{multline*}
        \mathbb{E}_{p_0}\biggl\{ \|\nabla_x \log p_0(X) -\nabla_x \log p_{\infty}(X) \|_2^2 \biggr\} \\
        = \min_{p \in \mathcal{A}_m} \mathbb{E}_{p} \biggl\{\|\nabla_x \log p(X) -\nabla_x \log p_{\infty}(X) \|_2^2 \biggr\}.
   \end{multline*}
\end{theorem}

\begin{proof}
    For any $p \in \mathcal{A}_m$, there exist $w_i$ such that $p = \sum_{i=1}^m w_i p_i$, where $w_i \geq 0$ and $\sum_{i=1}^m w_i = 1$. Direct calculations give
    \begin{align*}
        &\mathbb{E}_{p} \biggl\{ \|\nabla_x \log p(X) -\nabla_x \log p_{\infty}(X) \|_2^2 \biggr\}
        \\
        &= \mathbb{E}_{p}\biggl\{ \biggl\| \frac{\nabla_x p(X)}{ p(X)} -\nabla_x \log p_{\infty}(X)\biggr\|_2^2 \biggr\} 
         \\
        &= \mathbb{E}_{p}\biggl\{ \biggl\| \frac{\sum_{i=1}^m w_i \nabla_x p_i(X)}{ \sum_{i=1}^m w_i p_i(X)} -\nabla_x \log p_{\infty}(X) \biggr\|_2^2 \biggr\}  
        \\
        &= \mathbb{E}_{p}\biggl\{ \biggl\| \sum_{i=1}^m u_i(X) \nabla_x \log p_i(X) - \nabla_x \log p_{\infty}(X) \biggr\|_2^2 \biggr\},
    \end{align*}
    where $u_i(X) = \frac{w_ip_i(X)}{\sum_{i=1}^m w_ip_i(X)}$ for all $i=1,\ldots,m$, and $\sum_{i=1}^m u_i(X)=1$. Clearly $\nabla_x \log u_i(x) - \nabla_x \log u_j(x)  = \nabla_x \log  p_i(x) - \nabla_x \log  p_j(x)$ for all $1 \le i,j \le m$.
    
    Using Condition~(\ref{eq10}), the quantity above is minimized at $p = p_0$, which concludes the proof. 
\end{proof}
Theorem~\ref{thm_general_LFD} provides an efficient way to identify the LFD in a convex set with only knowledge of the gradient of the log density functions. 

Next, we provide a method to find the LFD in a class of Gaussian mixture models. 
\begin{theorem}
\label{theorem: lfd_example}
    Let $G_{\theta}$ denote the $d$-dimensional Gaussian distribution centered at $\theta \in \mathbb{R}^d$ with a constant covariance matrix $V \in \mathbb{R}^{d \times d}$. Let the set  $\Theta_1 \subseteq \mathbb{R}^d$ be compact and convex.
    Consider the pre-change distribution $G_{\theta_*}$ and post-change distribution class $\mathcal{G}_1$ defined as all Gaussian mixture models given by the convex hull of
    $\{ G_{\theta}: \theta \in \Theta_1 \}$.  
    For any vector $v \in \mathbb{R}^d$, let $\|v\|_V = (v^T V^{-2} v)^{1/2}$.
    Assume that $\theta_* \not\in \Theta_1$, and $\theta_0 \in \Theta_1$ is the closest to $\theta_*$ under the $\|\cdot\|_V$ norm, namely $\|\theta_0-\theta_*\|_V = \min_{\theta \in \Theta_1 }\|\theta-\theta_*\|_V $. Then, $G_{\theta_0}$ is the closest to $G_{\theta_*}$ among $\mathcal{G}_1$ under the Fisher divergence.
\end{theorem}

\begin{proof}
Let $g_{\theta_0}$ and $g_{\theta_*}$ denote the densities of $G_{\theta_0}$ and $G_{\theta_*}$, respectively.  Clearly,
   \begin{multline*}
   \min_{g_{\theta} \in \mathcal{G}_1 }\,  \mathbb{E}_{g_{\theta}} \biggl\{\| \nabla_x \log g_{\theta}(X) -\nabla_x \log g_{\theta_*}(X) \|_2^2 \biggr\} \\
\le 
\mathbb{E}_{g_{\theta_0}} \biggl\{\| \nabla_x \log g_{\theta_0}(X) -\nabla_x \log g_{\theta_*}(X) \|_2^2 \biggr\}
\end{multline*} 
We will prove the equality by proving the reverse inequality. To this end, consider an arbitrary element of $\mathcal{G}_1$. By definition of convex hull, this element can be written as $G_1 = \sum_{i=1}^m w_i G_{\theta_i}(X)$ for some $m \ge 1$,  $w_i \ge 0, i=1, \cdots, m$ with $\sum_{i=1}^{m} w_i = 1$ and $\theta_i \in \Theta_1$ for $i=1, \cdots, m$. As proved in the above theorem 
    \begin{align*}
        &\mathbb{E}_{g_1} \biggl\{ \|\nabla_x \log g_1(X) -\nabla_x \log g_{\theta_*}(X) \|_2^2 \biggr\}
        \\
        &= \mathbb{E}_{g_1}\biggl\{ \biggl\| \sum_{i=1}^m \beta_i(X) \nabla_x \log g_{\theta_i}(X) - \nabla_x \log g_{\theta_*}(X) \biggr\|_2^2 \biggr\},
    \end{align*}
    where $\beta_i(X) = \frac{w_ig_{\theta_i}(X)}{\sum_{i=1}^m w_i g_{\theta_i}(X)}$ for all $i=1,\ldots,m$.

Thus, we have
    \begin{align*}
& \mathbb{E}_{g_{1}} \biggl\{\| \nabla_x \log g_{1}(X) -\nabla_x \log g_{\theta_*}(X) \|_2^2 \biggr\} \\    
        &=  \mathbb{E}_{g_{1}}\biggl\| \sum_{i=1}^m \beta_i(X) (X - \theta_i) - (X - \theta_*) \biggr\|_V^2 \\
        &= \mathbb{E}_{g_{1}}\biggl\| \sum_{i=1}^m \beta_i(X) (\theta_* - \theta_i) \biggr\|_V^2.
    \end{align*}
    Using the assumption that $\|\theta_0-\theta_*\|_V = \min_{\theta \in \Theta_1 }\|\theta-\theta_*\|_V $, we have 
        \begin{align*}
        &= \mathbb{E}_{g_{1}}\biggl\| \sum_{i=1}^m \beta_i(X) (\theta_* - \theta_i) \biggr\|_V^2 \\
        &= \mathbb{E}_{g_{1}}\biggl\| \theta_* - \sum_{i=1}^m \beta_i(X) \theta_i \biggr\|_V^2  \geq  \mathbb{E}_{g_{\theta}}\| \theta_* - \theta_0 \|_V^2 \\
        &= 
        \mathbb{E}_{g_{\theta_0}} \biggl\{\| \nabla_x \log g_{\theta_0}(X) -\nabla_x \log g_{\theta_*}(X) \|_2^2 \biggr\}.
    \end{align*}
This concludes the proof.
\end{proof}

For a general parametric family of potential post-change distributions, it may be difficult to identify the LFD. In Section~\ref{subsec: example_lfd}, we propose a method to find the LFD in parameter space. 

\section{Numerical Results} \label{sec:results}
In this section, we present numerical results for both synthetic and real data demonstrating the robustness of RSCUSUM. 
Specifically, we identify the LFD in $\mathcal{G}_1$ defined as convex hull of given distributions $P_i(x), \, i=1,2, \cdots, m$. 
To this end,  we minimize the Fisher divergence over the set $\mathcal{B}_m$ defined in Equation (\ref{eq:convex_gradient_density}) and
invoke Theorem \ref{thm_general_LFD}.
In general, we can then estimate the $\nabla_x \log p_0(x)$ for LFD by $\sum_{i=1}^{m} \beta_i(x) \nabla_x \log p_i(x)$.  


\subsection{Example of the Least Favorable Distribution}
\label{subsec: example_lfd}
We consider the parametric family $\mathcal{P}$ as the multivariate Normal distribution (MVN), a subfamily~\citep{yu2016statistical} of the exponential family (EXP), and the Gauss-Bernoulli Restricted Boltzmann Machine (RBM)~\citep{LeCun2006ATO}.
For example, in the case of MVN, 
\begin{align*}
    &\mathcal{G}_{\infty} = \{\mathcal{N}(\mathbf{\mu}_*, V_*)\},\nonumber\\
    &\mathcal{G}_1= \left\{\sum_{i=1}^m \alpha_i\mathcal{N}(\boldsymbol{\mu}_i, V_i):\; \sum_{i=1}^m \alpha_i = 1, \;\forall \;\alpha_i\geq 0\right\}.
\end{align*}
Here the pre-change distribution $P_{\infty}=\mathcal{N}(\mathbf{\mu}_*, V_*)$ and the uncertainty class $\mathcal{G}_1$ is constructed from a finite basis $\mathcal{P}_m=\{\mathcal{N}(\boldsymbol{\mu}_i, V_i), \;i=1,\ldots, m\}$ (see Equation~(\ref{eq:convex_post_family})). Each basis element $P_i$ is parameterized by the corresponding vector $\boldsymbol{\theta}_i=(\boldsymbol{\mu}_i, V_i)$. Without loss of generality, we assume $\boldsymbol{\theta}_1$ to be the closest to $\boldsymbol{\theta}_*=(\boldsymbol{\mu}_{\star}, V_{\star})$ in $L_2$ (Euclidean) norm. 

By Theorem~\ref{thm_general_LFD}, it is sufficient to find $P_0$ such that Condition~\eqref{eq10} holds.  Any $\phi(x)\in\mathcal{B}_m$ is characterized by coefficients $\beta_j(\cdot), \;j=1,\ldots,m$ (see Equation~\eqref{eq:convex_gradient_density}).

We use a neural network $\operatorname{Softmax}_j\circ f_{\textit{NN}}(x)$ to estimate $\beta_j(\cdot)$, specifically, 
\begin{align*}
    \beta_j(x) = \operatorname{Softmax}_j\circ f_{\textit{NN}}(x),
\end{align*}
where $f_{\textit{NN}}$ is given by the feature extractor part of a multi-layer neural network corresponding to hidden layer sizes $[128-64-m]$, with $\operatorname{Softmax}$ the last layer
all $\operatorname{ReLU}$ activation functions in hidden layers.  Note that $\operatorname{Softmax}_j$ denotes the $j$-th element of the Softmax function.
The use of Softmax function ensures $\sum_{i=1}^m\beta_i(x)=1$ and $\beta_i(x) \ge 0$ for all $1 \le i \le m$.

To identify $P_0$, we learn $f_{\textit{NN}}$ by minimizing the following loss function over the training sample $X_1, \cdots, X_N\sim P$:
\begin{equation*} 
    \mathcal{L} = \frac{1}{N}\sum_{i=1}^N \biggl\|\sum_{j=1}^m\beta_j(X_i)\nabla\log p_{j}(X_i)-\nabla\log p_{\infty}(X_i)\biggr\|_2^2,
\end{equation*}
where $P$ is updated at each epoch based on the learned coefficients $\beta_i(x)$ by
$$\nabla_{x}\log p(x) = \sum_{i=1}^m \beta_{i}(x)\nabla_{x}\log p_{i}(x).$$
To generate samples from the unnormalized density function $\nabla_{x} \log p(x)$, standard Markov Chain Monte Carlo (MCMC)  techniques (such as MALA) are employed. Furthermore, the neural network is trained using the Adam optimizer.

In Table~\ref{tab: lfd_coeffs}, we report the average value $\frac{1}{M}\sum_{i=1}^M\beta_j(Y_i)$ over the test sample $Y_1, \cdots Y_M\sim P$ respectively in cases where the basis elements of $\mathcal{P}_m$ are MVN$_m$ (with mean shifts), MVN$_c$ (with covariance shifts), EXP, and RBMs. Details of $P_{\infty}$ and basis elements of $\mathcal{P}_m$ are given in the Supplementary Material. In all cases the average value of $\beta_1(y)$ (respectively $\beta_j(y), \,  j=2,3,4$) is extremely close to $1$ (respectively to $0$). This gives strong evidence that the LFD is achieved by one of the basis $\mathcal{P}_m$, and Theorem \ref{thm_general_LFD} can be invoked to give the LFD.

\begin{table}[htbp]
    \centering
    \begin{tabular}{c|cccc}
    \toprule
         j & $1$&$2$&$3$&$4$\\
         \hline
        MVN$_m$ &1.00e+00& 4.90e-09& 2.43e-11& 6.29e-12\\
        MVN$_c$ &9.99e-01& 7.47e-06& 3.23e-08& 3.55e-08\\
        EXP & 9.99e-01& 2.84e-05& 1.37e-09& 1.01e-09\\
        RBM & 1.00e+00& 3.18e-33& 0.00e+00& 0.00e+00\\
        \bottomrule
    \end{tabular}
    \caption{Empirical average values of $\beta_j(x)$ over $10000$ test sample for MVN, EXP, and RBM models.}
    \label{tab: lfd_coeffs}
\end{table}

\subsection{Synthetic Data} \label{subsec: synthetic_data}
As in Subsection~\ref{subsec: example_lfd}, we simulate synthetic data streams from MVNs and RBMs to evaluate the performance of RSCUSUM. The LFD in the uncertainty class is identified as in Subsection~\ref{subsec: example_lfd}. We also report the performance of the SCUSUM (which is not robust) \cite{wuetal-aistat-2023}  for arbitrary \textit{wrong} distributions in the uncertain class.

We consider a change detection scenario where the pre- and post-distributions are modeled by MVN (respectively RBM) models with $m=4$. Both $P_\infty$ and the elements of the uncertainty class are created according to detailed descriptions in the supplementary material. We use Gibbs sampling method with $1000$ iterations for RBMs. 
In each trial, we treat one of $P_i\in \mathcal{P}_m, \, i=1,2,3,4$ as the \textit{true} post-change distribution. For each trial, we perform the experiment for $1000$ runs. 
 
In all experiments, we set the change point as $\nu=50$, and we set the total length of each data stream as $10000$ to assure the generated data stream is long enough for detection.
We evaluate the detection delay for ARL values ranging from $100$ to $3000$. 

\begin{figure}[htbp]
\centering
 \includegraphics[width= 0.9\linewidth]{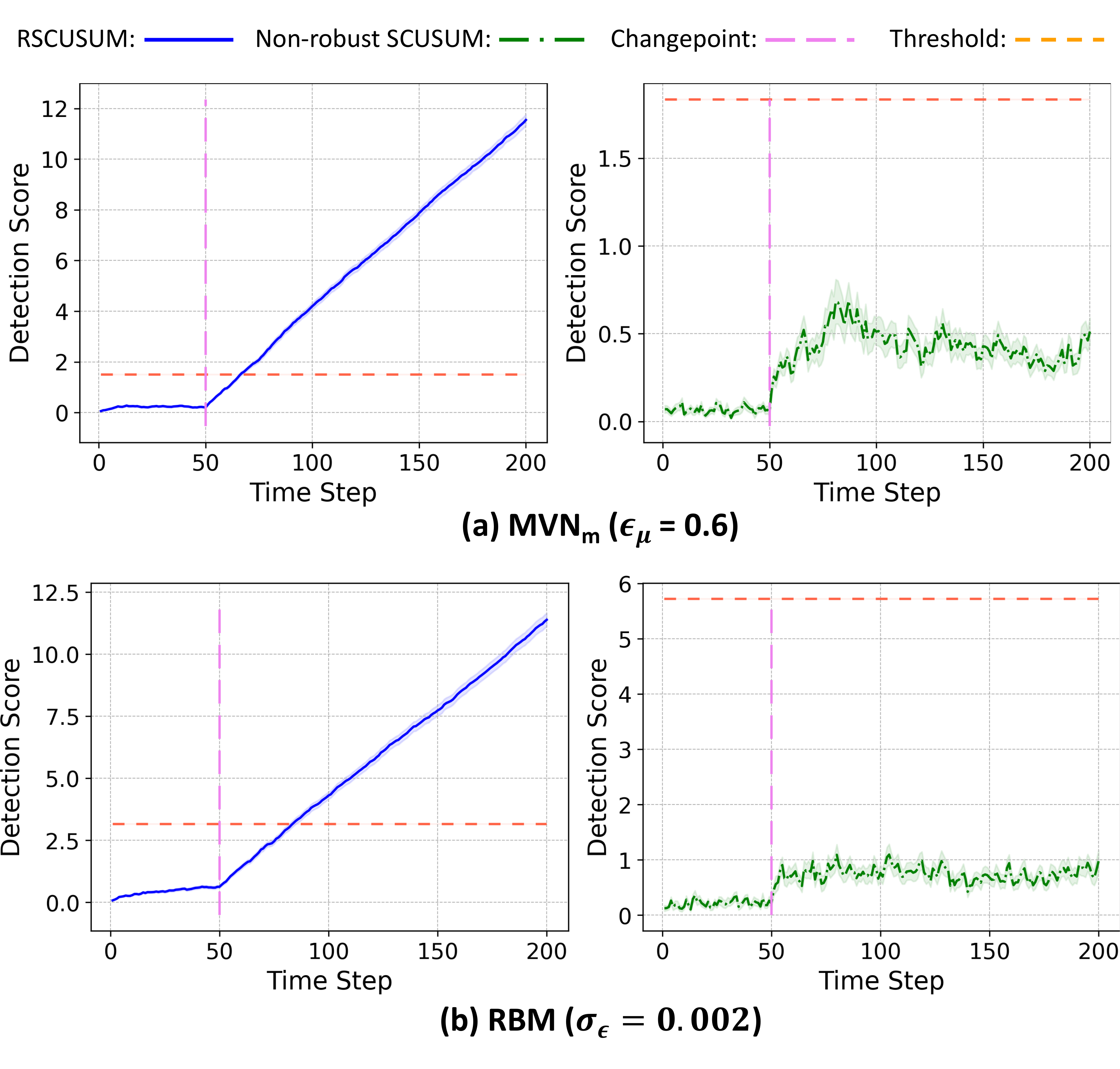}
   \caption{Detection score versus Time. }
 \label{fig:score}
\end{figure}

In Figure~\ref{fig:score}(a) and (b), we respectively report the detection scores versus time in cases for MVN$_m$ and RBM experiments. The results demonstrate that the average increment of detection scores is positive for RSUCUM, while negative for the non-robust SCUSUM. This means that a non-robust CUSUM fails to detect this post-change scenario but the RSCUSUM algorithms detects it. 

\begin{figure}[tbph]
\centering
 \includegraphics[width=0.9 \linewidth]{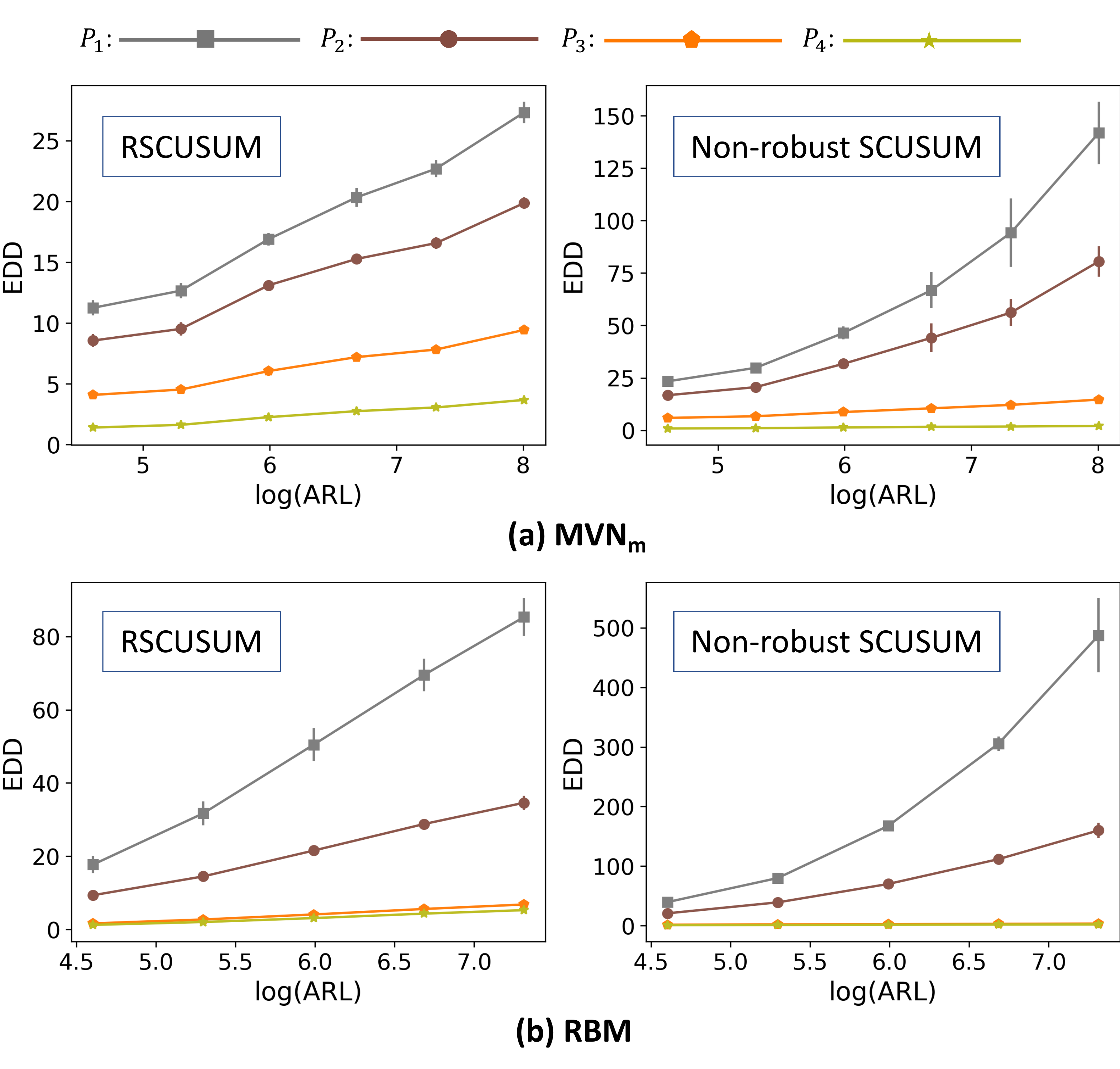}
   \caption{EDD versus log-scaled ARL. }
 \label{fig:mvn-mean}
\end{figure}

In Figure~\ref{fig:mvn-mean}(a) and (b), we respectively demonstrate the empirical EDD against log-scaled ARL for both MVN$_m$ and RBM experiments. The results demonstrate that  RSCUSUM is robust and performs competitively in terms of detection delay.  In particular, we observe that the EDD of RSCUSUM (subplot in left rows) increases at a linear rate for all cases, while some EDD of non-robust SCUSUM (subplot in right rows) may increase at an exponential rate (compare the y-axis labels for the plots).


\section{Conclusions} \label{sec: conclusion}
In this work, we proposed the RSCUSUM algorithm, a robust score-based algorithm for quickest change detection when the post-change distribution is not precisely known.  
We defined the least favorable distribution in the sense of Fisher divergence. Using asymptotic analysis, we also analyzed the delay and false alarms of RSCUSUM in the sense of Lorden's and Pollak's metrics. 
We provided both theoretical and algorithmic methods for computing the least favorable distribution for unnormalized models. Numerical simulations were provided to demonstrate the performance of our robust algorithm.

\begin{acknowledgements} 
Suya Wu and Vahid Tarokh were supported in part by Air Force Research Lab Award under grant number FA-8750-20-2-0504. Jie Ding was supported in part by the Office of Naval Research under grant number N00014-21-1-2590. Taposh Banerjee was supported in part by the U.S. Army Research Lab under grant W911NF2120295.
\end{acknowledgements}

\balance
\bibliography{uai2023-ref}

\begin{thebibliography}{22}
\providecommand{\natexlab}[1]{#1}
\providecommand{\url}[1]{\texttt{#1}}
\expandafter\ifx\csname urlstyle\endcsname\relax
  \providecommand{\doi}[1]{doi: #1}\else
  \providecommand{\doi}{doi: \begingroup \urlstyle{rm}\Url}\fi

\bibitem[Basseville et~al.(1993)Basseville, Nikiforov,
  et~al.]{basseville1993detection}
Michele Basseville, Igor~V Nikiforov, et~al.
\newblock \emph{Detection of abrupt changes: theory and application}, volume
  104.
\newblock prentice Hall Englewood Cliffs, 1993.

\bibitem[Doob(1953)]{doob1953stochastic}
Joseph~L Doob.
\newblock \emph{Stochastic processes}, volume~7.
\newblock Wiley New York, 1953.

\bibitem[Hyv{\"a}rinen(2005)]{hyvarinen2005estimation}
Aapo Hyv{\"a}rinen.
\newblock Estimation of non-normalized statistical models by score matching.
\newblock \emph{J. Mach. Learn. Res.}, 6\penalty0 (4), 2005.

\bibitem[Lai(1998)]{lai1998information}
Tze~Leung Lai.
\newblock Information bounds and quick detection of parameter changes in
  stochastic systems.
\newblock \emph{IEEE Trans. Inf. Theory}, 44\penalty0 (7):\penalty0 2917--2929,
  1998.

\bibitem[LeCun et~al.(2006)LeCun, Chopra, Hadsell, Ranzato, and
  Huang]{LeCun2006ATO}
Yann LeCun, Sumit Chopra, Raia Hadsell, M~Ranzato, and F~Huang.
\newblock A tutorial on energy-based learning.
\newblock In \emph{Predicting structured data}, volume~1. The MIT Press, 2006.

\bibitem[Lorden(1970)]{lorden1970excess}
Gary Lorden.
\newblock On excess over the boundary.
\newblock \emph{Ann. Math. Stat.}, 41\penalty0 (2):\penalty0 520--527, 1970.

\bibitem[Lorden(1971)]{lorden1971procedures}
Gary Lorden.
\newblock Procedures for reacting to a change in distribution.
\newblock \emph{Ann. Math. Stat.}, pages 1897--1908, 1971.

\bibitem[Moustakides(1986)]{moustakides1986optimal}
George~V Moustakides.
\newblock Optimal stopping times for detecting changes in distributions.
\newblock \emph{Ann. Stat.}, 14\penalty0 (4):\penalty0 1379--1387, 1986.

\bibitem[Page(1955)]{page1955test}
ES~Page.
\newblock A test for a change in a parameter occurring at an unknown point.
\newblock \emph{Biometrika}, 42\penalty0 (3/4):\penalty0 523--527, 1955.

\bibitem[Pollak(1985)]{pollak1985optimal}
Moshe Pollak.
\newblock Optimal detection of a change in distribution.
\newblock \emph{Ann. Stat.}, pages 206--227, 1985.

\bibitem[Poor and Hadjiliadis(2008)]{poor2008quickest}
H~Vincent Poor and Olympia Hadjiliadis.
\newblock \emph{Quickest detection}.
\newblock Cambridge University Press, 2008.

\bibitem[Roberts(1966)]{roberts1966comparison}
SW~Roberts.
\newblock A comparison of some control chart procedures.
\newblock \emph{Technometrics}, 8\penalty0 (3):\penalty0 411--430, 1966.

\bibitem[Shiryaev(1963)]{shiryaev1963optimum}
Albert~N Shiryaev.
\newblock On optimum methods in quickest detection problems.
\newblock \emph{Theory Probab. Appl.}, 8\penalty0 (1):\penalty0 22--46, 1963.

\bibitem[Song et~al.(2020)Song, Sohl-Dickstein, Kingma, Kumar, Ermon, and
  Poole]{song2020score}
Yang Song, Jascha Sohl-Dickstein, Diederik~P Kingma, Abhishek Kumar, Stefano
  Ermon, and Ben Poole.
\newblock Score-based generative modeling through stochastic differential
  equations.
\newblock \emph{arXiv preprint arXiv:2011.13456}, 2020.

\bibitem[Tartakovsky et~al.(2014)Tartakovsky, Nikiforov, and
  Basseville]{tartakovsky2014sequential}
Alexander Tartakovsky, Igor Nikiforov, and Michele Basseville.
\newblock \emph{Sequential analysis: Hypothesis testing and changepoint
  detection}.
\newblock CRC Press, 2014.

\bibitem[Tartakovsky and Veeravalli(2005)]{tartakovsky2005general}
Alexander~G Tartakovsky and Venugopal~V Veeravalli.
\newblock General asymptotic bayesian theory of quickest change detection.
\newblock \emph{Theory of Probability \& Its Applications}, 49\penalty0
  (3):\penalty0 458--497, 2005.

\bibitem[Unnikrishnan et~al.(2011)Unnikrishnan, Veeravalli, and
  Meyn]{unnikrishnan2011minimax}
Jayakrishnan Unnikrishnan, Venugopal~V Veeravalli, and Sean~P Meyn.
\newblock Minimax robust quickest change detection.
\newblock \emph{IEEE Transactions on Information Theory}, 57\penalty0
  (3):\penalty0 1604--1614, 2011.

\bibitem[Veeravalli and Banerjee(2014)]{veeravalli2014quickest}
Venugopal~V Veeravalli and Taposh Banerjee.
\newblock Quickest change detection.
\newblock In \emph{Academic press library in signal processing}, volume~3,
  pages 209--255. Elsevier, 2014.

\bibitem[Woodroofe(1982)]{woodroofe1982nonlinear}
Michael Woodroofe.
\newblock \emph{Nonlinear renewal theory in sequential analysis}.
\newblock SIAM, 1982.

\bibitem[Wu et~al.(2023)Wu, Diao, Banerjee, Ding, and
  Tarokh]{wuetal-aistat-2023}
Suya Wu, Enmao Diao, Taposh Banerjee, Jie Ding, and Vahid Tarokh.
\newblock Score-based change point detection for unnormalized models.
\newblock \emph{International Conference on Artificial Intelligence and
  Statistics (AISTATS)}, 2023.

\bibitem[Xie et~al.(2021)Xie, Zou, Xie, and Veeravalli]{xie2021sequential}
Liyan Xie, Shaofeng Zou, Yao Xie, and Venugopal~V Veeravalli.
\newblock Sequential (quickest) change detection: Classical results and new
  directions.
\newblock \emph{IEEE Journal on Selected Areas in Information Theory (JSAIT)},
  2\penalty0 (2):\penalty0 494--514, 2021.

\bibitem[Yu et~al.(2016)Yu, Kolar, and Gupta]{yu2016statistical}
Ming Yu, Mladen Kolar, and Varun Gupta.
\newblock Statistical inference for pairwise graphical models using score
  matching.
\newblock \emph{Advances in Neural Information Processing Systems (NeurIPS)},
  29, 2016.

\end{thebibliography}

\clearpage

\title{Robust Quickest Change Detection for Unnormalized Models\\(Supplementary Material)}

%
%
  
\onecolumn 
\maketitle
\appendix
\section{Likelihood Ratio-based Robust CUSUM Algorithm}
\label{subsec:llr-cusum}
\noindent In this section, we review the result in \cite{unnikrishnan2011minimax} on classical robust quickest change detection. Let $p_{\infty}$ and $p_{1}$ be the density functions of pre- and post-change distributions. If the post-change law is known, then given the data stream $\{X_n\}_{n\geq 1}$, the stopping rule of the likelihood ratio-based CUSUM algorithm is defined by
\begin{equation} 
\label{eq:cusumrule}
    T_{\texttt{CUSUM}}=\inf\{n\geq 1:\Lambda(n)\geq \tau\},
\end{equation}
where $\Lambda(n)$ is defined using the recursion
\begin{align}
    &\Lambda(0)=0, \nonumber \\
    &\Lambda(n) \de \biggr(\Lambda(n-1)+\log \frac{p_1(X_n)}{p_{\infty}(X_n)}\biggr)^{+}, \forall n \geq 1, \label{eq:cusum_score1}
 \end{align}
which leads to a computationally efficient stopping scheme (if the densities $p_1$ and $p_{\infty}$ are precisely known). 
In \cite{moustakides1986optimal}, it is shown that the CUSUM algorithm is exactly optimal, for every fixed constraint $\gamma$, for Lorden's problem. As pointed out in \cite{lai1998information}, the algorithm is also asymptotically optimal for Pollak's problem. In \cite{lorden1971procedures} and \cite{lai1998information}, the asymptotic performance of the CUSUM algorithm is also characterized. Specifically, it is shown as $\gamma \rightarrow \infty$.
\begin{align}
\label{eq:optimality_cusum}
    \mathcal{L_{\texttt{WADD}}}(T_{\texttt{CUSUM}}) \sim \mathcal{L_{\texttt{CADD}}}(T_{\texttt{CUSUM}})\sim \frac{\log \gamma}{\mathbb{D}_{\texttt{KL}}(P_{1}\|P_{\infty})}.
\end{align}
Here $\mathbb{D}_{\texttt{KL}}(p_{1}\|p_{\infty})$ is the Kullback-Leibler divergence between the post-change density $p_1$) and pre-change distribution $p_{\infty}$:
$$
\mathbb{D}_{\texttt{KL}}(P_{1}\|P_{\infty}) = \int_x p_1(x) \log \frac{p_1(x)}{p_\infty(x)} dx, 
$$
and the notation $g(c)\sim h(c)$ as $c\to c_0$ indicates that $\frac{g(c)}{h(c)} \to 1$ as $c\to c_0$ for any two functions $c\mapsto g(c)$ and $c\mapsto h(c)$.

The CUSUM algorithm can successfully detect a change in law from $p_1$ to $p_\infty$ because 
\begin{equation}
\label{eq:driftCUSUM}
    \begin{split}
        \int_x &\log \frac{p_1(x}{p_\infty(x)} p_1(x) dx = \mathbb{D}_{\texttt{KL}}(P_{1}\|P_{\infty}) > 0 \\
         \int_x &\log \frac{p_1(x}{p_\infty(x)} p_\infty(x) dx = -\mathbb{D}_{\texttt{KL}}(P_\infty\|P_1) < 0.
    \end{split}
\end{equation}
Thus, the mean of the increment of $\Lambda(n)$
in \eqref{eq:cusum_score1} before the change is negative, and after the change is positive. 

If the post-change density $p_1$ is not known and assumed to belong to a family $\mathcal{G}_1$, then the test is designed using the least favorable distribution. Specifically, in \cite{unnikrishnan2011minimax}, it is assumed that there is a density $q_1 \in \mathcal{G}_1$ such that for every $p_1 \in \mathcal{G}_1$, 
\begin{equation}
\label{eq:leastfavunni}
    \begin{split}
        \log \frac{q_1(X)}{p_\infty(X)} \bigg|_{X \sim q_1} \; \; \prec \quad \; \; \log \frac{q_1(X)}{p_\infty(X)}\bigg|_{X \sim p_1} . 
    \end{split}
\end{equation}
Here the notation $\prec$ is used to denote stochastic dominance: if $W$ and $Y$ are two random variables, then $W \prec Y$ if
$$
P(Y \geq t) \geq P(W \geq t), \quad \text{for all } t \in (-\infty, \infty). 
$$
If such a density $q_1$ exists in the post-change family, then the robust CUSUM is defined as the CUSUM test with $q_1$ used as the post-change density. Such a test is exactly optimal for the problem of \cite{lorden1971procedures} under additional assumptions on the smoothness of densities, and asymptotically optimal for the problem in \cite{pollak1985optimal}. We refer the reader to \cite{unnikrishnan2011minimax} for a more precise optimality statement. 

We note that in the literature on quickest change detection, the issue of the unknown post-change model has also been addressed by using a generalized likelihood ratio (GLR) test or a mixture-based test. While these tests have strong optimality properties, they are computationally even more expensive than the robust test described above; see \cite{lorden1971procedures, lai1998information, tartakovsky2014sequential}. 

As discussed in the introduction, the robust CUSUM algorithm discussed above may have two major drawbacks: 1) Due to the complicated characterization of the least favorable distribution $q_1$ \eqref{eq:leastfavunni}, it may be hard to identify in high-dimensional models. 2) The robust CUSUM is a likelihood ratio-based test and is thus computationally expensive to implement for high-dimensional models. 

In Section 4 of the main paper, we propose the RSCUSUM algorithm to mitigate these issues. 
\begin{enumerate}
    \item The RSCUSUM algorithm is based on Hyv\"arinen score (\cite{hyvarinen2005estimation}) and is invariant to normalizing constants. This makes it computationally efficient for high-dimensional models which are often only learnable within a normalizing constant. 
    \item We defined the notion of least favorable distribution differently in our paper. For us, the least favorable distribution has the least Fisher divergence with respect to the pre-change model. We also provided an efficient computational method to identify the least favorable distribution. 
\end{enumerate}

\section{Proofs}
The theoretical analysis for delay and false alarms is analogous to that of analysis from \cite{wuetal-aistat-2023}. We give complete proofs here for completeness.

\subsection{Proof of Lemma 4.3}
\begin{proof}
Define the function $\lambda:\mapsto h(\lambda)$ given by $$h(\lambda)\de\mathbb{E}_{\infty}[\exp (z_{\lambda}(X))]-1.$$ Observe that \begin{equation*}
  h^{\prime}(\lambda)\de \frac{d h}{d\lambda}(\lambda)=\mathbb{E}_{\infty}[(S_{\texttt{H}}(X,P_{\infty})-S_{\texttt{H}}(X,Q_{1}))\exp (z_{\lambda}(X))].
\end{equation*}
Note that $h(0)=0$, and $h^{\prime}(0)=-\mathbb{D}_{\texttt{F}}(P_{\infty}\|Q_1)<0$. 
Next, we prove that either 1) there exists $\lambda^{\star} \in (0,\infty)$ such that $h(\lambda^{\star}) = 0$, or 2) for all $\lambda>0$ we have $h(\lambda)<0$. 

Observe that
\begin{equation*}
    h''(\lambda)\de\frac{d^2 h}{d \lambda}(\lambda)
    =\mathbb{E}_{\infty}[(S_{\texttt{H}}(X,P_{\infty})-S_{\texttt{H}}(X,Q_{1}))^2\exp (z_{\lambda}(X))]\geq 0.
\end{equation*}
We claim that $h(\lambda)$ is \textit{strictly convex}, namely $h''(\lambda) > 0$ for all $\lambda\in [0,\infty)$. Suppose $h''(\lambda) = 0$ for some $\lambda \geq 0$, we must have $S_{\texttt{H}}(X,P_{\infty})-S_{\texttt{H}}(X,Q_1) = 0$ almost surely.  This implies that 
$\mathbb{E}_{\infty}[(S_{\texttt{H}}(X,P_{\infty})-S_{\texttt{H}}(X,Q_1))]
= 0$ which in turn gives $-\mathbb{D}_{\texttt{F}}(P_{\infty}\|Q_1) =0$ and $P_{\infty} = Q_1$ almost everywhere, leading to a contradiction to the assumption $P_{\infty}\notin \mathcal{G}_1$. Thus, $h(\lambda)$ is \textit{strictly convex} and $h^{\prime}(\lambda)$ is \textit{strictly increasing}. 

Here, we recognize two cases: either 1) $h(\lambda)$ have at most one global minimum in $(0, \infty)$, or 2) it is strictly decreasing in $[0,\infty)$. We will show that the second case is degenerate that is of no practical interest.
\begin{itemize}
\item \textbf{Case 1:} If the global minimum of $h(\lambda)$ is attained at $a \in (0, \infty)$, then $h^{\prime}(a) = 0$. Since $h^{\prime}(0) < 0$ and $h(0) = 0$, the global minimum $h(a)<0$. 
Since $h^{\prime}(\lambda)$ is \textit{strictly increasing}, we can choose $b > a$ and conclude that $h^{\prime}(\lambda) > h^{\prime}(b) > h^{\prime}(a) = 0$ for all $\lambda > b$. It follows that $\lim_{\lambda \rightarrow \infty} h(\lambda) = +\infty$. Combining this with the continuity of $h(\lambda)$, we conclude that $h(\lambda^*) = 0$ for some $\lambda^* \in (0, \infty)$ and any value of $\lambda \in (0, \lambda^*]$ satisfies Inequality~(10).

Note that in this case, we must have $P_{\infty}\left(S_{\texttt{H}}(X,P_{\infty})-S_{\texttt{H}}(X,Q_1) \ge c\right)>0$, for some $c>0$. Otherwise, we have $P_{\infty}\left(S_{\texttt{H}}(X,P_{\infty})-S_{\texttt{H}}(X,Q_1) \le 0\right)=1$. This implies that $P_{\infty}(z_{\lambda}(X)\le 0)=1$, or equivalently $\mathbb{E}_{\infty}[\exp (z_{\lambda}(X))]< 1$ for all $\lambda > 0$, and therefore leads to Case 2: $h(\lambda)< 0$ for all $\lambda > 0$. Here, $\mathbb{E}_{\infty}[\exp (z_{\lambda}(X))]\neq 1$ since $P_{\infty}(S_{\texttt{H}}(X,P_{\infty})-S_{\texttt{H}}(X,Q_1)=0)<1$; otherwise $P_{\infty}(S_{\texttt{H}}(X,P_{\infty})-S_{\texttt{H}}(X,Q_1)=0)=1$, and then $\mathbb{E}_{\infty}[S_{\texttt{H}}(X,P_{\infty})-S_{\texttt{H}}(X,Q_1)]=-\mathbb{D}_{\texttt{F}}(P_{\infty}\|Q_1)=0$, causing the same contradiction to $P_{\infty}\notin\mathcal{G}_1$.

\item \textbf{Case 2:} If $h(\lambda)$ is strictly decreasing in $(0, \infty)$, then any $\lambda \in (0, \infty)$ satisfies Inequality~(10). As discussed before, in this case, we must have $P_{\infty}\left(S_{\texttt{H}}(X, P_{\infty})-S_{\texttt{H}}(X, Q_1) \le 0\right)=1$. Equivalently, all the increments of the RSCUSUM detection score are non-positive under the pre-change distribution, and $P_{\infty}(Z(n)=0)=1$ for all $n$. Accordingly, $\mathbb{E}_{\infty}[T_{\textit{RSCUSUM}}]=+\infty$. When there occurs change (under measure $Q_1$), we also observe that RSCUSUM can get close to detecting the change point instantaneously as $\lambda$ is chosen arbitrarily large. Obviously, this case is of no practical interest.
\end{itemize}

\end{proof}

\subsection{Proof of Theorem 4.4}
\begin{proof}
We follow the proof of \cite{lai1998information}[Theorem 4] to conclude the result of Theorem 4.4. A constructed martingale and Doob's submartingale inequality~\citep{doob1953stochastic} are combined to finish the proof. 
\begin{enumerate}
    \item We first construct a non-negative martingale with mean $1$ under the measure $P_{\infty}$. Define a new instantaneous score function $X \mapsto \tilde{z}_{\lambda}(X)$ given by 
\begin{equation*}
    \label{eq:new_instant_z_lambad}
    \tilde{z}_{\lambda}(X)\de z_{\lambda}(X)+\delta,
\end{equation*}
where $$\delta \de -\log \biggr(\mathbb{E}_{\infty}\left[\exp (z_{\lambda}(X))\right]\biggr).$$ Further define the sequence $$\tilde{G}_n\de \exp \biggr(\sum_{k=1}^n\tilde{z}_{\lambda}(X_k)\biggr),\; \forall n\geq 1.$$ 

Suppose $X_1, X_2, \ldots$ are i.i.d according to $P_{\infty}$ (no change occurs). Then,
\begin{align*}
\mathbb{E}_{\infty}\left[\tilde{G}_{n+1}\mid \mathcal{F}_n\right] = \tilde{G}_n\mathbb{E}_{\infty}[\exp(\tilde{z}_{\lambda}(X_{n+1}))]=\tilde{G}_{n}e^{\delta}\mathbb{E}_{\infty}[\exp(z_{\lambda}(X_{n+1}))]=\tilde{G}_{n},
\end{align*}
and
\begin{align*}
    \mathbb{E}_{\infty}[\tilde{G}_n] &= \mathbb{E}_{\infty}\left[\exp\left(\sum_{i=1}^{n}(z_{\lambda}(X_i)+\delta)\right)\right]= e^{n\delta} \prod_{i=1}^n\mathbb{E}_{\infty}[\exp(z_{\lambda}(X_i))]=1.
\end{align*}
Thus, under the measure $P_{\infty}$, $\{\tilde{G}_n\}_{n\geq 1}$ is a non-negative martingale with the mean $\mathbb{E}_{\infty}[\tilde{G}_1]=1$. 

\item We next examine the new stopping rule 
\begin{equation*}
    \tilde{T}_{\texttt{RSCUSUM}} = \inf \left\{n\geq 1: \max_{1\leq k\leq n} \sum_{i=k}^n \tilde{z}_{\lambda}(X_i)\geq \tau \right\},
\end{equation*}
where $\tilde{z}_{\lambda}(X_i) = z_{\lambda}(X_i)+\delta$. By Inequality~(10), we observe that $\delta\geq 0$. By Jensen's inequality,
\begin{equation}
\label{eq:jensen}
    \mathbb{E}_{\infty}[\exp(z_{\lambda}(X))]\geq \exp\left(\mathbb{E}_{\infty}[z_{\lambda}(X)]\right),
\end{equation}
with equality holds if and only if $z_{\lambda}(X)=c$ almost surely, where $c$ is some constant. Suppose the equality of Equation~(\ref{eq:jensen}) holds, then\begin{align*}
    -\lambda \mathbb{D}_{\texttt{F}}(Q_1||P_{\infty})&=\mathbb{E}_{\infty}[z_{\lambda}(X)]=c=\mathbb{E}_{1}[z_{\lambda}(X)]=\lambda \mathbb{D}_{\texttt{F}}(P_{\infty}||Q_1).
\end{align*} 
It follows that $0\leq \mathbb{D}_{\texttt{F}}(P_{\infty}||Q_1)=-\mathbb{D}_{\texttt{F}}(Q_1||P_{\infty})\leq 0$, which implies that $P_{\infty}\notin \mathcal{G}_1$ almost everywhere. This leads to a contradiction to the assumption $P_{\infty}\notin \mathcal{G}_1$. Thus, the inequality of Equation~(\ref{eq:jensen}) is \textit{strict}, and therefore $\delta<\lambda\mathbb{D}_{\texttt{F}}(P_{\infty}||Q_1)$. Hence, $\tilde{T}_{\texttt{RSCUSUM}}$ is not trivial.

Define a sequence of stopping times: 
\begin{align*}
    &\eta_0 = 0,\\
    &\eta_1 = \inf \left\{t:\sum_{i=1}^t \tilde{z}_{\lambda}(X_i)<0\right\},\\
    &\eta_{k+1} = \inf \left\{t>\eta_k:\sum_{i=\eta_k+1}^t \tilde{z}_{\lambda}(X_i)<0\right\}, \; \text{for}\;  k\geq 1.
\end{align*}
By previous discussion, $\{\tilde{G}_n\}_{n\geq 1}$ is a nonnegative martingale under $P_{\infty}$ with mean 1. Then, for any $k$ and on $\{\eta_k<\infty\}$,
\begin{equation}
\label{eq:doobs}
P_{\infty}\left(\sum_{i=\eta_k+1}^n\tilde{z}_{\lambda}(X_i)\geq \tau \;\text{for some}\;  n>\eta_k \mid \mathcal{F}_{\eta_k} \right) \leq e^{-\tau},
\end{equation}
by Doob's submartingale inequality~\citep{doob1953stochastic}. Let
\begin{equation}
\label{eq:defm}
    M \de \inf \biggl\{k\geq 0: \eta_k<\infty \;\text{and} \; \sum_{i=\eta_k+1}^n\tilde{z}_{\lambda}(X_i)\geq \tau \; \text{for some}\; n>\eta_k\biggr\}.
\end{equation}
Combining Inequality~(\ref{eq:doobs}) and Definition~(\ref{eq:defm}),
\begin{align}
\label{eq:eq2}
    P_{\infty}(M\geq k+1\mid\mathcal{F}_{\eta_k})= 1-P_{\infty}\left(\sum_{i=\eta_k+1}^n\tilde{z}(X_i)\geq \tau  \;\text{for some} \; n>\eta_k\mid \mathcal{F}_{\eta_k}\right)\geq 1-e^{-\tau},
\end{align}
and
\begin{equation}
\label{eq:eq1}
    P_{\infty}(M> k)= \mathbb{E}_{\infty} [P_{\infty}(M\geq k+1\mid\mathcal{F}_{\eta_k})\mathbb{I}_{\{M\geq k\}}]=\mathbb{E}_{\infty}[P_{\infty}(M\geq k+1\mid\mathcal{F}_{\eta_k})]P_{\infty}(M> k-1).
\end{equation}
Combining Equations~(\ref{eq:eq1}) and (\ref{eq:eq2}), 
\begin{align*}
    \mathbb{E}_{\infty}[M] = \sum_{k=0}^{\infty}P_{\infty}(M> k)\geq \sum_{k=0}^{\infty}(1-e^{-\tau})^{k}= e^{\tau}.
\end{align*}

Observe that
\begin{align*}
    \tilde{T}_{\texttt{RSCUSUM}}&=\inf \biggl\{n\geq 1:\sum_{i=\eta_k+1}^n\tilde{z}_{\lambda}(X_i)\geq \tau \; \text{for some}\; \eta_k<n \biggr\}\geq M,
\end{align*}
and $\tilde{T}_{\texttt{RSCUSUM}}\leq T_{\texttt{RSCUSUM}}$. We conclude that
$\mathbb{E}_{\infty}[T_{\texttt{RSCUSUM}}]\geq \mathbb{E}_{\infty}[\tilde{T}_{\texttt{RSCUSUM}}]\geq \mathbb{E}_{\infty}[M]\geq e^{\tau}$.
\end{enumerate}
\end{proof}

\subsection{Proof of Theorem 4.5}
We first introduce a technical definition in order to apply~\cite{woodroofe1982nonlinear}[Corollary 2.2.] to the proof of Theorem 4.5.
\begin{definition}
A distribution $P$ on the Borel sets of $(-\infty, \infty)$ is said to be \textit{arithmetic} if and only if it concentrates on a set of points of the form $\pm nd$, where $d>0$ and $n=1, 2, \ldots$.
\end{definition}
\begin{remark}
    Any probability measure that is absolutely continuous with respect to the Lebesgue measure is non-arithmetic.
\end{remark}
\begin{proof}
Consider the random walk that is defined by 
\begin{equation*}
    Z^{\prime}(n) = \sum_{i=1}^nz_{\lambda}(X_i), \; \text{for}\; n\geq 1.
\end{equation*}
We examine another stopping time that is given by
\begin{equation*}
     T_{\texttt{RSCUSUM}}^{\prime} \de \inf \{n\geq 1: Z^{\prime}(n) \geq \tau\}.
\end{equation*}
Next, for any $\tau$, define $R_{\tau}$ on $\{T_{\texttt{RSCUSUM}}^{\prime} <\infty\}$ by 
\begin{equation*}
    R_{\tau} \de Z^{\prime}(T_{\texttt{RSCUSUM}}^{\prime}) -\tau.
\end{equation*}
$R_{\tau}$ is the excess of the random walk over a stopping threshold $\tau$ at the stopping time $T_{\texttt{RSCUSUM}}^{\prime}$.
Suppose the change point $\nu =1$, then $X_1, X_2,\ldots, $ are i.i.d. following the distribution $Q_1$. Let $\mu$ and $\sigma^2$ respectively denote the mean $\mathbb{E}_{1}[z_{\lambda}(X)]$ and the variance $\text{Var}_1[z_{\lambda}(X)]$. Note that 
\begin{equation*}
    \mu =\mathbb{E}_{1}[z_{\lambda}(X)]= \lambda(\mathbb{D}_{\texttt{F}}(P_1\|P_{\infty})-\mathbb{D}_{\texttt{F}}(P_1\|Q_{1}))>0,
\end{equation*}
and \begin{equation*}
    \sigma^2 = \text{Var}_1[z_{\lambda}(X)] = \mathbb{E}_1[z_{\lambda}(X)^2]-\left(\lambda(\mathbb{D}_{\texttt{F}}(P_1\|P_{\infty})-\mathbb{D}_{\texttt{F}}(P_1\|Q_{1}))\right)^2.
\end{equation*}
Under the mild regularity conditions
given by \cite{hyvarinen2005estimation},
\begin{align*}
&\mathbb{E}_{1}[\mathcal{S}_{\texttt{H}}(X, P_{\infty})]^2 < \infty,\;\text{and} \\
&\mathbb{E}_{1}[\mathcal{S}_{\texttt{H}}(X, Q_1)]^2 < \infty.
\end{align*}
It implies that $\mathbb{E}_1[z_{\lambda}(X)^2]<\infty$ if $\lambda$ is chosen appropriately, e.g. $\lambda$ satisfy Inequality~(14) and $\lambda$ is not arbitrary large. 
Therefore, by \cite{lorden1970excess} Theorem 1,
\begin{equation*}
    \sup_{\tau \geq 0}\mathbb{E}_1[R_{\tau}]\leq \frac{\mathbb{E}_1[(z_{\lambda}(X)^{+})^2]}{\mathbb{E}_1[z_{\lambda}(X)]}\leq \frac{\mu^2+\sigma^2}{\mu},
\end{equation*}
where $z_{\lambda}(X)^{+} = \max (z_{\lambda}(X), 0)$.
Additionally,  $Q_1$ must be non-arithmetic in order to have Hyv\"arinen scores well-defined. Hence, by \cite{woodroofe1982nonlinear} Corollary 2.2.,
\begin{equation*}
    \mathbb{E}_{1}[T^{\prime}_{\texttt{RSCUSUM}}]=\frac{\tau}{\mu}+\frac{\mathbb{E}_1[{R_{\tau}}]}{\mu}\leq \frac{\tau}{\mu}+\frac{\mu^2+\sigma^2}{\mu^2},\;\forall \tau \geq 0.
\end{equation*}
Observe that for any $n$, $Z^{\prime}(n)\leq Z(n)$, and therefore $T_{\texttt{RSCUSUM}} \leq T_{\texttt{RSCUSUM}}^{\prime}$. Thus, 
\begin{equation}
\label{eq:cadd_result}
    \mathbb{E}_{1}[T_{\texttt{RSCUSUM}}]\leq \mathbb{E}_{1}[T_{\texttt{RSCUSUM}}^{\prime}]\leq \frac{\tau}{\mu}+\frac{\mu^2+\sigma^2}{\mu^2},\;\forall \tau \geq 0.
\end{equation}
By Theorem 4, we select $\tau = \log \gamma $ to satisfy the constraint $\mathbb{E}_{\infty}[T_{\texttt{RSCUSUM}}]\geq\gamma>0$. Plugging it back to Equation~(\ref{eq:cadd_result}), we conclude that, as $\gamma \to \infty$,
\begin{equation*}
    \mathbb{E}_{1}[T_{\texttt{RSCUSUM}}] \sim \frac{\log \gamma}{\mu}=\frac{\log \gamma}{\lambda(\mathbb{D}_{\texttt{F}}(P_1\|P_{\infty})-\mathbb{D}_{\texttt{F}}(P_1\|Q_{1}))},
\end{equation*}
to complete the proof.

Due to the stopping scheme of RSCUSUM, the expected time $\mathbb{E}_{\nu}[T_{\texttt{RSCUSUM}}-\nu|T_{\texttt{RSCUSUM}}\geq \nu]$ is independent of the change point $\nu$ (This is obvious, and the same property for CUSUM has been shown by~\cite{xie2021sequential}). Let $\nu = 1$, and we have \begin{equation*}
    \mathcal{L}_{\texttt{CADD}}(T_{\texttt{RSCUSUM}}) = \mathbb{E}_{1}[T_{\texttt{RSCUSUM}}]-1.
\end{equation*}
Thus, we conclude that 
\begin{equation*}
    \mathcal{L}_{\texttt{CADD}}(T_{\texttt{RSCUSUM}})\sim \frac{\log \gamma}{\lambda (\mathbb{D}_{\texttt{F}}(P_1\|P_{\infty})-\mathbb{D}_{\texttt{F}}(P_1\|Q_{1}))}.
\end{equation*}
Similar arguments applies for $\mathcal{L}_{\texttt{WADD}}(T_{\texttt{RSCUSUM}})$.
\end{proof}

\subsection{Selection of Appropriate Multiplier}

It is worth noting that although results of our core results hold for a pre-selected $\lambda$ that satisfied the condition discussed in Lemma 4.3. The effect of choosing any other $\lambda^{\prime}$ amounts to the scaling of all the increments of RSCUSUM by a constant factor of $\lambda^{\prime}/ \lambda$. This means that all of these results still hold adjusted for this scale factor. For instance, the result of Theorem 4.4 can be modified to be written as 
$$
\mathbb{E}_{\infty}[T_{\texttt{RSCUSUM}}]\geq \exp \left\{\frac{\lambda  \tau}{\max(\lambda, \lambda^{\prime})}\right\},
$$ 
for any $\lambda^{\prime} > 0$. It is easy to see that this scaling will change the statement of Theorem 4.5 accordingly to 
$$
\mathbb{E}_{1}[T_{\texttt{RSCUSUM}}]\sim \frac{\max(\lambda, \lambda^{\prime})}{\lambda }\frac{\log \gamma}{\lambda^{\prime}(\mathbb{D}_{\texttt{F}}(P_1\|P_{\infty})-\mathbb{D}_{\texttt{F}}(P_1\|Q_{1}))},
$$ 
as $\gamma \to \infty$. In order to have the strongest results in Theorems 4.4 and 4.5, we must choose $\lambda$ as close to $\lambda^*$ as possible.

\section{Experimental Details}
\subsection{Synthetic Dataset}
We consider the parametric family $\mathcal{P} = \{G_{\theta}:\;\theta\in\Theta\}$, and a set of basis elements $\mathcal{P}_m=\{P_1,\ldots, P_m\}$, $\forall P_i\in \mathcal{P}$. We set $m=4$ for synthetic simulations. The uncertainty class of post-change distribution (pre-change distribution respectively) is given by
\begin{align*}
    &\mathcal{G}_1= \left\{\sum_{i=1}^m \alpha_iP_i:\; \sum_{i=1}^m \alpha_i = 1, \;\forall \;\alpha_i\geq 0\right\},\nonumber\\
    &\mathcal{G}_{\infty} = \{P_{\infty}:\: P_{\infty}\in\mathcal{P},\;P_{\infty}\notin\mathcal{G}_1\}.
\end{align*}

\paragraph{Multivariate Normal Distribution (MVN)} Let $\boldsymbol{\mu}$ and $V$ respectively denote the mean and the covariance matrix. The corresponding score function is calculated by
\begin{equation*}
    S_{\texttt{H}}(X, P) = \frac{1}{2}(X-\boldsymbol{\mu})^{T}\Sigma^{-2}(X-\boldsymbol{\mu})-\operatorname{tr}(V^{-1}),
\end{equation*}
where the operator $\operatorname{tr}(\cdot)$ takes the trace of matrix.

For the scenario of MVN$_m$, we think the covariance matrix $V$ is a constant for any distribution in the parametric family. The pre-change distribution $P_{\infty}=\mathcal{N}(\mathbf{\mu}_*, V_*)$, where \begin{align*}
    \boldsymbol{\mu}_{\star}=(0,0),\quad\text{and}\quad V_{\star} = \left(\begin{matrix}
    1, &0.5\\
    0.5, &1
\end{matrix}\right).
\end{align*}
The set $\mathcal{P}_m=\{\mathcal{N}(\boldsymbol{\mu}_j, V_j), \;j=1,\ldots, m\}$, where 
\begin{align*}
    \boldsymbol{\mu}_j=(\epsilon_j,\epsilon_j),\quad\text{and}\quad V_{j} = \left(\begin{matrix}
    1, &0.5\\
    0.5, &1
\end{matrix}\right).
\end{align*}
We take the value of $\epsilon_1$ ($\epsilon_j$, $j=2,3,4$ respectively) as $0.5$ ($0.6, 0.8, 1.0$ respectively) for $P_1$ ($P_j$, $j=2,3,4$ respectively).

For the scenario of MVN$_c$, we consider both the mean and covariance matrix as the parameter. Again, we consider the pre-change distribution $P_{\infty}=\mathcal{N}(\mathbf{\mu}_*, V_*)$, and the set $\mathcal{P}_m=\{\mathcal{N}(\boldsymbol{\mu}_j, V_i), \;j=1,\ldots, m\}$. Here,  
\begin{align*}
    \boldsymbol{\mu}_j=(\epsilon_j,\epsilon_j),\quad\text{and}\quad V_{j} = \left(\begin{matrix}
    1, &0.5\\
    0.5, &1
\end{matrix}\right)\circ\exp(\delta_j),
\end{align*}
where $\circ$ denotes the element-wise product and $\epsilon_{\log(\sigma^2)}$ denotes the element-wise perturbations of the covariance matrix. We take the value of $\delta_j$ (respectively $\delta_j, j=2,3,4$) as $0.1$ ($0.2, 0.8, 1.0$ respectively) for $P_1$ ($P_j$, $j=2,3,4$ respectively). To make the perturbed covariance matrix positive-definite, we perturb the log of each component of the covariance matrix.

\begin{table}[ht]
\centering
\caption{EDD versus ARL for RSCUSUM and RCUSUM on Multivariate Gaussian Case}
\begin{tabular}{ c| c c c c c c c }
\toprule
Perturbation/ARL &  & 100 & 200 & 400 & 800 & 1500 & 3000 \\
\hline
\multirow{2}{*}{0.5} & RSCUSUM & 11.2552 & 12.6664 & 16.9057 & 20.3400 & 22.7026 & 27.3190 \\
& RCUSUM & 11.4017 & 12.8748 & 16.8437 & 20.2776 & 22.6781 & 27.2831 \\
\hline
\multirow{2}{*}{0.6} & RSCUSUM & 8.5636 & 9.5218 & 13.1102 & 15.2747 & 16.5815 & 19.8648 \\
& RCUSUM & 8.6460 & 9.5817 & 12.9797 & 15.2196 & 16.5526 & 19.7900 \\
\hline
\multirow{2}{*}{1} & RSCUSUM & 4.0894 & 4.5327 & 6.0542 & 7.1984 & 7.8237 & 9.4318 \\
& RCUSUM & 4.1259 & 4.5658 & 6.0447 & 7.1551 &7.8026 & 9.3947 \\
\hline
\multirow{2}{*}{2} & RSCUSUM & 1.4053 & 1.6268 & 2.2620 & 2.7546 & 3.0592 & 3.6752 \\
& RCUSUM & 1.4290 & 1.6393 & 2.2516 & 2.7393 & 3.0481 & 3.6684 \\
\bottomrule
\end{tabular}
\end{table}

\paragraph{Exponential Family (EXP)} We consider the Exponential family with the associated PDF given by
\begin{align}
    p_{\theta}(X) =\frac{1}{Z_{\tau}} \exp\left\{-\tau\left(\sum_{i=1}^d(x_i-\mu)^4+\sum_{1\leq i\leq d, i\leq j\leq d}(x_i-\mu)^2(x_j-\mu)^2\right)\right\},\nonumber
\end{align}
where $\theta = (\tau, \mu)$.
The associated Hyvarinen score function is calculated by
\begin{equation*}
    S_{\texttt{H}}(X, P_{\theta}) = \frac{1}{2}\sum_{i=1}^d \left(\frac{\partial}{\partial x_i}\log P_{\theta}(X)\right)^2+\sum_{i=1}^d\frac{\partial^2}{\partial x_i}\log P_{\theta}(X),
\end{equation*}
where 
\begin{align*}
    \frac{\partial}{\partial x_i}\log P_{\theta}(X) &= -\tau \left(4(x_i-\mu)^3+2\sum_{1\leq i\leq d, i\leq j\leq d}(x_i-\mu)(x_j-\mu)^2\right), \;\text{and}\\
    \frac{\partial^2}{\partial x_i}\log P_{\theta}(X)&=-\tau \left (12(x_i-\mu)^2+2\sum_{1\leq i\leq d, i\leq j\leq d}(x_j-\mu)^2\right).
\end{align*}
We consider the pre-change distribution $P_{\infty}$ with $\tau_{\star} =1$ and $\mu_*=0$. The post-change distribution basis elements are constructed with $\tau =\tau_*+\epsilon_{j}$ and $\mu=\mu_*+\delta_j$. Here, $\epsilon_{j}$ ($\delta_j$ respectively) denotes the perturbations of the scale parameter $\tau$ (the location parameter $\mu$ respectively) for each $P_j$, $j=1,2,3,4$. We take values of $\epsilon_{j}$ as $1.0, 2.0, 8.0, 10.0$, and values of $\delta_j$ as $0.01, 0.02, 0.08, 0.1$.

\paragraph{Gauss-Bernoulli Restricted Boltzmann Machine (RBM)} As introduced in Subsection the main paper, we consider the RBM mode with the PDF given by $p_{\theta}(X)= \sum_{h\in \{0,1\}^{d_h}}p_{\theta}(X, H) = \frac{1}{Z_{\theta}}\exp\{-F_{\theta}(X)\}$, where $F_{\theta}(X)$ is the free energy given by 
\begin{equation*}
    F_{\theta}(X) = \frac{1}{2}\sum_{i=1}^{d_x} (x_{i}-b_i)^{2}\nonumber
    -\sum_{j=1}^{d_h} \operatorname{Softplus}\left(\sum_{i=1}^{d_x} W_{i j}x_{i}+b_{j}\right).
\end{equation*}
We compute the corresponding Hyv\"arinen score in a closed form
\begin{equation*}
    S_{\texttt{H}}(X, P_{\theta})= \sum_{i=1}^{d_x}\left[\frac{1}{2}\left(x_{i}-b_{i}+\sum_{j=1}^{d_h} W_{ij} \phi_{j}\right)^2+\sum_{j=1}^{d_h} W_{i j}^{2} \phi_{j}\left(1-\phi_{j}\right)-1\right],
\end{equation*}
where $\phi_{j} \de \operatorname{Sigmoid}(\sum_{i=1}^{d_x} W_{i j}x_{i}+b_{j})$. The $\operatorname{Sigmoid}$ function is defined as $\operatorname{Sigmoid}(y) \de (1+\exp(-y))^{-1}$.

The pre-change distribution $P_{\infty}$ is with the parameters $\mathbf{W} = \mathbf{W}_*$, $\mathbf{b}=\mathbf{b}_*$, and $\mathbf{c}=\mathbf{c}_*$, where each component of $\mathbf{W}_*$, $\mathbf{b}_*$, and $\mathbf{c}_*$ is randomly drawn from the standard Normal distribution $\mathcal{N}(0,1)$. For the post-change distribution basis elements, we assign the parameters $\mathbf{W}_j=\mathbf{W}_*\oplus\epsilon_j$, $\mathbf{b}_j=\mathbf{b}_*$, and $\mathbf{c}_j=\mathbf{c}_*$. Here, we only consider shifts of weight matrix $\mathbf{W}$. We let $\epsilon_j$ take values from $0.001, 0.002, 0.008, 0.01$ for $P_j$, $j=1,2,3,4$.


\end{document}